\documentclass[english,prl,twocolumn,superscriptaddress,amsmath,amssymb,a4paper]{revtex4}
\usepackage[T1]{fontenc}
\usepackage[latin1]{inputenc}
\usepackage{graphicx}
\usepackage[top=0.85in,bottom=0.9in,,left=0.75in,right=0.75in]{geometry}
\usepackage{booktabs}
\usepackage{color}
\usepackage{verbatim}
\usepackage{enumitem}

\makeatletter
\usepackage{dcolumn}
\usepackage{bm}
\usepackage{epsfig}

\newcommand{\ignore}[1]{}
\usepackage{appendix}
\usepackage{babel}
\makeatother

\usepackage{sidecap}
\sidecaptionvpos{figure}{c}

\begin{document}

\title{Ferromagnetic Negative Charge-Transfer Insulator: from Theoretical Proposal to Material Realization}

\author{Zhao Liu}
\affiliation{Hefei National Laboratory for Physical Sciences at the Microscale, University of Science and Technology of China, Hefei, Anhui 230026, China}

\author{Xingxing Li}
\affiliation{Hefei National Laboratory for Physical Sciences at the Microscale,
Synergetic Innovation Center of Quantum Information and Quantum Physics,
University of Science and Technology of China, Hefei, Anhui 230026, China}

\author{W. Zhu}
\affiliation{Institute of Natural Sciences, Westlake Institution of Advanced Study and
School of Science, Westlake University, Hangzhou 310024, China}

\author{Z. F. Wang} 
\affiliation{Hefei National Laboratory for Physical Sciences at the Microscale,
CAS Key Laboratory of Strongly-Coupled Quantum Matter Physics,
University of Science and Technology of China, Hefei, Anhui 230026, China}

\author{Jinlong Yang} \thanks{E-mail: jlyang@ustc.edu.cn}
\affiliation{Hefei National Laboratory for Physical Sciences at the Microscale,
Synergetic Innovation Center of Quantum Information and Quantum Physics,
University of Science and Technology of China, Hefei, Anhui 230026, China}


\begin{abstract}
Here we propose another type of ferromagnetic semiconductors: ferromagnetic negative charge-transfer insulator (FNCTI). 
In FNCTI, the negative charge-transfer states strongly enhance the ferromagnetic (FM) exchange interactions and the orbital hybridization gap permits the magnetic molecular orbitals as the underlying magnetic units rather than local atomic orbitals. Thus the FM exchange interactions are rather strong and decay slowly due to the large spearding of magnetic molecular orbitals.
This is distinct from the superexchange mechanism where FM exchange  interactions are quite weak as summarized in the well-known Goodenough-Kanamori-Anderson semi-empirical rules. 
Through first-principle calculations with the hybrid functional, PbO-type CrAs monolayer is mapped out to be a FNCTI, which possesses a band gap $\sim$ 0.35 eV, FM nearest-/next-nearest-neighbor exchange coupling strength $\sim$ 57/40 meV, and a high $T_c$ $\sim$ 1500 K respectively. It is believed that the existence of FNCTI validates the long-pending hypothesis by D. I. Khomskii and G. A. Sawatzky in 1997 [{\color{blue}Solid State Commun. \textbf{102}, 87 (1997)}].

\end{abstract}

\maketitle

\section{I. Introduction}
Ferromagnetism (FM), one of the oldest but most mysterious phenomena, is still intriguing intensive studies \cite{Brando2016, Hasegawa2009, Meng2018, Chen2021, Jin2022}. The existence/vanishing of charge gap with FM gives itinerant FM/FM semiconductor (or insulator). Different mechanisms have been proposed for itinerant FM, like the early Nagaoka's theorem \cite{Nagaoka1966, Tasaki1989}, flat-band ferromagnetism \cite{Mielke1991_1, Mielke1991_2, Tasaki1992, Mielke1993, Hase2018, Wang2022}, multiorbital system \cite{Shen1989, Li2014} etc. For ferromagnetic semiconductor (FMSC), the underlying mechanism varies for intrinsic and extrinsic system. The extrinsic system, also known as diluted FMSC, is obtained via doping magnetic ions into the nonmagnetic intrinsic semiconductor \cite{DFM1, DFM2}. The multiple degree of freedoms, including charge, spin, orbital and impurity, make it hard to write down a unified theory \cite{DFM3, DFM4, DFM5, Chen2021}. At the same time, the difficulty in manipulating magnetic impurities greatly hinders their developments \cite{Dietl2014}. Herein we will focus on intrinsic FMSC.

The intrinsic combination between FM and gap can trace back to the idea of superexchange interaction, first proposed by H. A. Kramers \cite{Kramers1934} and then developed by P. W. Anderson \cite{Anderson1950}. In contrast with the itinerant FM where direct exchange between correlated orbitals are possible, superexchange interaction relies on ligand $p$ orbitals to mediate long-range exchange interactions. Later on several quantitative relations on FM and AFM superexchange interaction were unveiled, mainly by J. B. Goodenough \cite{Goodenough1955, Goodenough1958}, J. Kanamori \cite{Kanamori1959} and P. W. Anderson \cite{Anderson1959}.

\begin{figure}
\centering
\includegraphics[width=6cm]{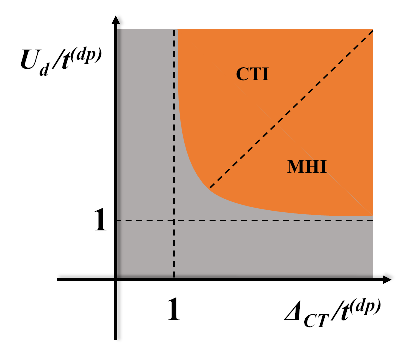}
\caption{Schematic illustration of Zaanen-Sawatzky-Allen scheme. The grey (orange) region is metallic (insulating) phase. }
\label{fig:ZSA}
\end{figure}

In the 1990s, J. Zaanen, G. A. Sawatzky and J. W. Allen solved the Anderson impurity model \cite{ZSA1985, ZSA1990} and classified all gapped transition metal compounds into two types: charge-transfer insulator (CTI) and Mott-Hubbard insulator (MHI) (see {\color{blue} Fig. \ref{fig:ZSA}}). These two insulators are indistinguishable at the ground state level (labelled as $|d^{n}L>$ or $|d^{n}>$), but they have different low energy excitations. Among the two characteristic one-particle excitations in transition metal compounds: $\Delta_{CT} = E(|d^{n+1}\underline{L}>) - E|d^{n}L>$ ($\underline{L}$ means a hole on $L$) which describes charge transfer from $p$ to $d$ and $U_d = E(|d^{n+1}d^{n-1}>) - E(|d^{n}d^{n}>)$ which characterizes charge fluctuation inner $d$ shell, we have $\Delta_{CT}$ < $U_d$ in CTI  while $\Delta_{CT}$ > $U_d$ in MHI.
These low energy excitations give rise to exchange interaction (with strength labelled by $J$) when the fermionic model is reduced to spin model. 
Since the condition $t^{(dp)} << \{\Delta_{CT}, U_d \}$ (where $t^{(dp)}$ is the hopping strength between $d$ and $p$ orbitals) is satisfied in both CTI and MHI, a perturbative treatment of kinetic energy is possible. Under such circumstances, the exchange interactions can be derived in a unified form for both CTI and MHI. The obtained semi-empirical rules thus cover the aforementioned quantitative relations by J. B. Goodenough \cite{Goodenough1955, Goodenough1958}, J. Kanamori \cite{Kanamori1959} and P. W. Anderson \cite{Anderson1959} and are summarized as Goodenough-Kanomari-Anderson(GKA) semi-empirical rules nowadays \cite{Khomskii2014}, which lays the foundation of modern-day understanding of superexchange interaction. However GKA semi-empirical rules impose strong constraints on FM superexchange coupling strength in both CTI and MHI: the antiferromagnetic (AFM) is generally much stronger than FM superexchange interaction and FMSCs seldom have ambient Curie temperature ($T_c$), like the long-knowing bulk EuS (17 K) \cite{EuS-1967}, EuO (69 K) \cite{EuO-1961} and recently discovered 2D CrI$_3$ (45 K) \cite{CrI3-2017}, Cr$_2$Ge$_2$Te$_6$ (30 K) \cite{CrGeTe3-2017}.
Deviations from the standard $90^{\circ}$ and $180^{\circ}$ $d-p-d$ geometries are often seen in transition metal compounds. If the deviation is small, the above GKA semi-empirical rules are still available. When the deviation is large, more and more exchange channels are possible, GKA semi-empirical rules are less predictive. In this situation, first-principles calculation is a powerful tool to determine the competition. For materials with irregular $d-p-d$ angles between $90^{\circ}$ and $180^{\circ}$, the AFM (FM) $J$ is weakened (strengthened) with reducing $d-p-d$ angle, therefore it is possible to find a balance point with strong FM \cite{Huang2019}.

As room-temperature FMSCs are the core unit of next-generation spintronic devices, such as processing-in-memory, spin field-effect transistors, magnetic tunnelling junctions and so on \cite{spintronics1, spintronics2, spintronics3}, an alternative mechanism which can break the inborn bottleneck in the superexchange mechanism is thus highly needed. Nevertheless it seems that there is no other insulator phase besides MHI or CTI in {\color{blue} Fig. \ref{fig:ZSA}}. Fortunately, this is not the case and in recent years, a hidden insulating phase is discovered in the previously believed metallic phase of {\color{blue} Fig. \ref{fig:ZSA}} and is now known as negative CTI \cite{CrO2-1998, SrCoO3-2012, NaCuO2-1991, RNiO3-2016}. Since the charge gap is negative, holes are self-doped to the ligand $p$ orbitals even at the ground state level \cite{CrO2-1998}. It is believed that there is strong AFM direct exchange interaction between metal cations and ligand $p$ holes, therefore D. I. Khomskii and G. A. Sawatzky guessed a strong emergent FM $J$ between two metal cations in such kind of insulators in 1997 \cite{Khomskii1997} (see {\color{blue}Fig. \ref{fig:emFM}(a)}. Although been suggested for a long time, less improvements have been made along this line \cite{Li2018} when comparing with the well-established super-exchange mechanism. 

In this work, we give a systematic investigation on this idea which motivate us to propose the concept of ferromagnetic negative charge-transfer insulator (FNCTI). This is a type of FMSC with physical properties different from that of FM CTI or MHI. This paper is organized as follows: in section II, we set the stage by introducing the multi-band Hamiltonian to describe transition metal compounds. After a perturbative treatment on $90^{\circ}$ and $180^{\circ}$ $d-p-d$ cluster models and a mean-field treatment on the bipartite square lattice, the basic features of FNCTI are sketched. In section III, we focus on the the material realization of FNCTI. In principle, FNCTI can be realized in any lattice structure by varying related parameters. Here we take PbO-type monolayer as a prototype. By choosing Cr (As) as the transition metal (ligand), CrAs monolayer is found to be an ideal FNCTI. In the closure section IV, a detailed comparison between FNCTI and FM CTI (MHI) is made. What's more, other aspects of FNCTI are also discussed there. 

\begin{figure*}
\centering
\includegraphics[width=16cm]{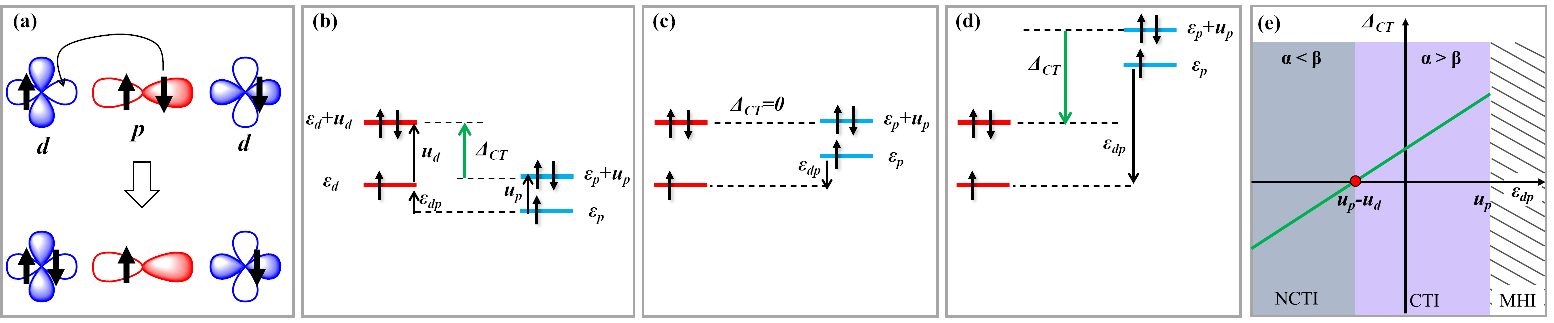}
\caption{(a) Real space description of $\Delta_{CT}$ in a cluster model formed by  a $p$ connecting two $d$ orbitals. (b)-(d) Energy space description of $\Delta_{CT}$ for $u_d$ > $\Delta_{CT}$ > 0, $\Delta_{CT}$ = 0 and $\Delta_{CT}$ < 0 for model in (a). Here only $\varepsilon_{p}$ is the variable with all other parameters fixed. (e) A schematic phase diagram for $\Delta_{CT}$ versus $\varepsilon_{dp}$.}
\label{fig:CTG}
\end{figure*}

\section{II. Hamiltonian for transition metal compounds}
Transition metal compounds are described by the multi-band $d-p$ model \cite{Mizokawa2000}:
\begin{equation}
\hat{H} = \hat{H}_d + \hat{H}_p + \hat{H}_{dp}  \\
\label{eq:Full-H}
\end{equation}
where $\hat{H}_d$ ($\hat{H}_p$) describes $d$ ($p$) shell of metal cations (ligand anions) and $\hat{H}_{dp}$ is the inter-shell term between $d$ and $p$ shells. For simplicity, we will use "M" and "$L$" for metal and ligand from now on. Each component in {\color{blue}Eq-(\ref{eq:Full-H})} is contributed by kinetic and interaction terms. For example, $\hat{H}_d$ is given by:
\begin{equation}
\begin{split}
\hat{H}_d &= \varepsilon_d \sum_{i, m, \sigma} \hat{n}^{(d)}_{im\sigma} +
\sum_{i, j, m, n, \sigma} (t^{(d)}_{im,jn} \hat{d}^{\dagger}_{im\sigma}
\hat{d}_{jn\sigma}+\textit{h.c.}) \\
& + u_{d} \sum_{i, m} \sum_{i, m}\hat{n}^{(d)}_{im\uparrow}\hat{n}^{(d)}_{im\downarrow} + u'_{d} \sum_{i, m \neq n}\hat{n}^{(d)}_{im\uparrow}\hat{n}^{(d)}_{in\downarrow} \\
&+  \frac{u'_d - j^{(d)}_H}{2}\sum_{m \neq n, \sigma}\hat{n}^{(d)}_{im\sigma}\hat{n}^{(d)}_{in\sigma} \\
&- j^{(d)}_H \sum_{m \neq n} (\hat{d}^{\dagger}_{im\uparrow}\hat{d}_{im\downarrow}\hat{d}^{\dagger}_{in\downarrow}\hat{d}_{in\uparrow} - \hat{d}^{\dagger}_{im\uparrow}\hat{d}^{\dagger}_{im\downarrow}\hat{d}_{in\downarrow}\hat{d}_{in\uparrow}) \\
\end{split}
\end{equation}
where i, m, $\sigma$ are indices for the site, orbital and spin degree of freedom, $\hat{d}^{\dagger}_{im\sigma}$ ($\hat{d}_{im\sigma}$) is the creation (annihilation) operator for a $d$ electron labelled by site i, orbital m and spin $\sigma$. 
The first term is the onsite energy ($\varepsilon^{(d)}$) of $d$ orbitals. The second term is the hopping energy ($t^{(d)}$) between different $d$ orbitals at different site, which gives $d$ bands. The left terms describes the intra-atomic Coulomb interactions expressed by Kanamori parameters $u_d$, $u'_d$ and $j^{(d)}_H$. To retain the rotational invariance in real space, we have the constraint: $u'_d = u_d - 2j^{(d)}_H$.

As for $p$ shell, $\hat{H}_p$ have a similar form as $\hat{H}_d$, and the corresponding creation (annihilation) operator is $\hat{p}^{\dagger}_{im\sigma}$ ($\hat{p}_{im\sigma}$), with the parameters $\varepsilon^{(p)}$, $t^{(p)}$, $u_p$, $u'_p$ and $j^{(p)}_H$. These intra-atomic Coulomb interactions are usually ignored, but in fact, these terms are not small at all, especially the Hund's coupling for $p$ (O: 1.2 eV) can be even larger than $d$ orbitals.

The last component is generally taken as hopping energy and inter-shell density-density interaction:
\begin{equation}
\begin{split}
\hat{H}_{dp} &= \sum_{i,j, m, n, \sigma} (t^{(dp)}_{im,jn}  \hat{d}^{\dagger}_{im\sigma} \hat{p}_{jn\sigma}+\textit{h.c.}) \\
&+ u_{dp} \sum_{<i,j>, m, n, \sigma, \sigma'} \hat{n}^{(d)}_{im\sigma} \hat{n}^{(p)}_{jn\sigma'}  \\
\end{split}
\label{eq:Hd}
\end{equation} 
where $t^{(dp)}$ is the orbital hybridization between $d$ and $p$ shell which is responsible for crystal field splitting and $u_{dp}$ describes the non-local inter-shell Coulomb interaction. Believed to be small, $u_{dp}$ is always abandoned so only $p-d$ hopping are considered. However, such an interaction is important in stabilizing charge gap in cuprates \cite{Emery1987, Hansmann2014} and modifying exchange interactions \cite{Eder1996}.  

The large parameter space makes the exact solution of {\color{blue}Eq-(\ref{eq:Full-H})} intractable and approximations are necessary.
For example, $p$ shell always lies below $d$ shell, thus it is reasonable to treat $d$ shell as the only active shell. Under such circumstance, it is reasonable to ignore $\hat{H}_p$, $\hat{H}_{dp}$ in {\color{blue}Eq-(\ref{eq:Full-H})} and the full model is simplified to the Anderson lattice model (ALM) with screened $u_d$, $u'_d$ and $j^{(d)}_H$. What's more, if the $d$ bands are narrow, it is safe to ignore $t^{(d)}$ and ALM is further reduced to Anderson lattice model which has been solved by J. Zaanen et al. as mentioned above \cite{ZSA1985, ZSA1990}. However when $p$ and $d$ orbitals are both active, the above assumption fails and both $p$ and $d$ orbitals should be treated as correlated. 

\begin{figure*}
\centering
\includegraphics[width=16cm]{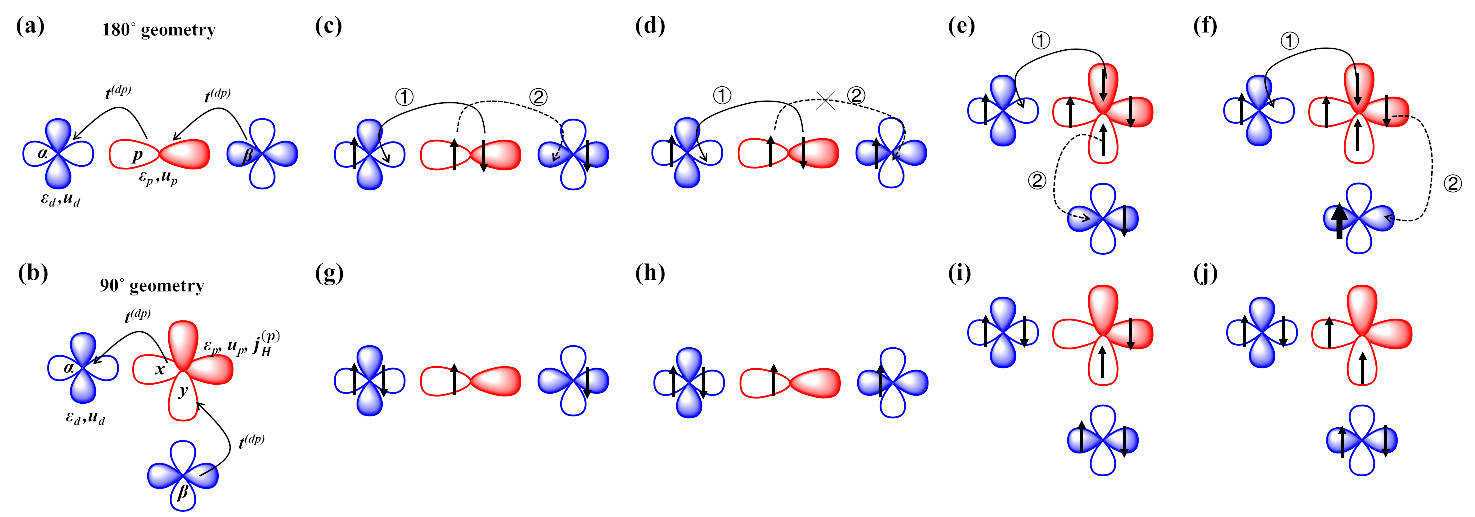}
\caption{(a) 180$^{\circ}$ $d-p-d$ geometry and (b) 90$^{\circ}$ geometry used in cluster model. Zero-th order state of (c) $S$ = 0 sector for 180$^{\circ}$ with $\Delta_{CT} >> 0$  (d) $S$ = 1 sector for 180$^{\circ}$ with $\Delta_{CT} >> 0$  (e) $S$ = 0 sector for 90$^{\circ}$ with $\Delta_{CT} >> 0$  (f) $S$ = 1 sector for 90$^{\circ}$ with $\Delta_{CT} >> 0$  (g) $S$ = 0 sector for 180$^{\circ}$ with $-u_p$ < $\Delta_{CT} << 0$  (h) $S$ = 1 sector for 180$^{\circ}$ with $\Delta_{CT} << 0$  (i) $S$ = 0 sector for 90$^{\circ}$ with $\Delta_{CT} << 0$  (j) $S$ = 1 sector for 90$^{\circ}$ with $\Delta_{CT} << 0$. The labels \textcircled{1} and \textcircled{2} represent the virtual electron hopping process between $d$ and $p$ orbital.}
\label{fig:cluster}
\end{figure*}

\begin{table*}[t]
	\caption{Results of $J$ for 180$^{\circ}$ and 90$^{\circ}$ $d-p-d$ geometry for CTI and NCTI at extreme condition, here $\Delta_{CT}$ is defined by {\color{blue}Eq.(\ref{eq:CT1})}}
	\label{tab:ana-J}
	\begin{tabular}{c|c|c}
		\hline\hline
		Geometry  & $\Delta_{CT} >> 0$  & $\Delta_{CT} << 0$  \\
		\hline
		180$^{\circ}$ & $\frac{4(t^{(dp)})^4}{\Delta_{CT}^2}$($\frac{1}{u_d}$ + $\frac{2}{2\Delta_{CT} + u_p}$) ($\Delta_{CT}$ < $u_d$) &  $\frac{4(t^{(dp)})^2}{\Delta_{CT} + u_p}$ ($\Delta_{CT} > - u_p$) \\
		90$^{\circ}$ & $-\frac{4(t^{(dp)})^4}{(\Delta_{CT}-2u_p+5j^{(p)}_H)^2}$ [$\frac{1}{(2\Delta_{CT}-3u_p+8j^{(p)}_H)-j^{(p)}_H}$ - $\frac{1}{(2\Delta_{CT}-3u_p+8j^{(p)}_H)+j^{(p)}_H}$ ] & $-2j^{(p)}_H$\\
		\hline\hline
	\end{tabular}
\end{table*}

\subsection{A: Negative charge-transfer insulator}
Since $\Delta_{CT}$ describes an electron hopping from $p$ to $d$ orbitals, it is a function of $\varepsilon_{d}$, $\varepsilon_{p}$ and the interaction parameters:
\begin{equation}
\Delta_{CT} = \varepsilon_{dp} + E_{int}(u_d, j^{(d)}_H, u_p, j^{(p)}_H)
\end{equation}
where $\varepsilon_{dp}$ is the onsite energy difference between $d$ and $p$ orbitals: $\varepsilon_{dp} = \varepsilon_{d} - \varepsilon_{p}$ and $E_{int}(u_d, j^{(d)}_H, u_p, j^{(p)}_H)$ describes the interaction contribution to $\Delta_{CT}$. A general expression of $E_{int}(u_d, j^{(d)}_H, u_p, j^{(p)}_H)$ is impossible as it depends on the $d$ and $p$ fillings. For the electron hopping process represented in {\color{blue} Fig. \ref{fig:CTG}(a)}, we have $E_{int}(u_d, j^{(d)}_H, u_p, j^{(p)}_H) = u_d - u_p$ as $p$ orbital is no longer fully occupied and $d$ orbital becomes fully occupied (seen in {\color{blue} Fig. \ref{fig:CTG}(b)}) so 
\begin{equation}
\Delta_{CT} = \varepsilon_{dp} + u_d - u_p
\label{eq:CT1}
\end{equation}
In most transition metal compounds, $p$ orbitals are below $d$ orbitals and if $\varepsilon_{dp} < u_p$, the system is a CHI as depicted in {\color{blue} Fig. \ref{fig:CTG}(b)}. Suppose we can tune $\varepsilon_{p}$ to higher energy (with $\varepsilon_{d}$ fixed), a special point is that $\varepsilon_{dp}$ = $u_p - u_d$ (or $\Delta_{CT}$ = 0) as shown in {\color{blue} Fig. \ref{fig:CTG}(c)}. At this point, since the transfer of the electron from $p$ to $d$ orbital doesn't consume energy, state $|d^{1}L>$ and $|d^{2}\underline{L}>$ have the same energy and the system is gapless. If $\varepsilon_{p}$ continues to increase, then $\Delta_{CT}$ becomes negative and the system becomes negative charge-transfer insulator (NCTI) (see {\color{blue} Fig. \ref{fig:CTG}(d)}). Configuration interaction tells us the ground state should be a linear combination of $|d^{1}L>$ and $|d^{2}\underline{L}>$ when $t^{(dp)}$ is included:
\begin{equation}
|GS> = \alpha |d^{1}L> + \beta |d^{2}\underline{L}>
\label{eq:mbs}
\end{equation}
For CTI/NCTI, because $|d^{1}L>$ has lower (higher) energy than $|d^{2}\underline{L}>$, $\alpha$ should be larger (smaller) than $\beta$ and they coincide with each other when $\Delta_{CT}$ = 0. There results are summarized in {\color{blue} Fig. \ref{fig:CTG}(e)}.
The fact that $|d^{2}\underline{L}>$ has lower energy than $|d^{1}L>$ in NCTI has huge impact on the magnetic exchange coupling. To see this point, we will use two different methods: a perturbative treatment of cluster models and a mean-field treatment of bipartite square lattice model.

\subsection{B: Perturbative treatment of cluster models}
Here we study the standard 90$^{\circ}$ and 180$^{\circ}$ $d-p-d$ geometry shown in {\color{blue} Fig. \ref{fig:cluster}(a)-(b)}, the parameters marked there are inherited from {\color{blue}Eq-(\ref{eq:Full-H})}. 

It is well known that Hubbard model preserves both U(1) charge and SU(2) spin symmetry, with electron number given, all the states can be further classified by the spin quantum number $S$. For Hubbard model with $S$ = 0 and $S$ = 1 sectors available, it is convenient to define $J$ as the energy difference:
\begin{equation}
J = E^{(0)}_{S = 1} - E^{(0)}_{S = 0}
\end{equation}
where $E^{(0)}_{S = 1}$, $E^{(0)}_{S = 0}$ are the energy of ground state in the $S$ = 1 and $S$ = 0 sector. So $J$ < 0 (> 0) indicates a FM (AFM) exchange coupling. 

Here we consider two extremes: $\Delta_{CT} >> 0$ (but still in the CTI regime) and $\Delta_{CT} << 0$ (but avoiding fully empty $p$ orbital) so a perturbative treatment is possible. {\color{blue}Tab. (\ref{tab:ana-J})} lists the main results and the derivation can be found in {\color{blue}Appendix B}. From {\color{blue}Tab. (\ref{tab:ana-J})}, it is clear that the exchange coupling mechanism has different behaviour at these two extremes. For $\Delta_{CT} >> 0 $, $J$ of both 180$^{\circ}$ and 90$^{\circ}$ cases are quartic power of $t^{(dp)}$ and 180$^{\circ}$ gives strong AFM while 90$^{\circ}$ gives weak FM, reflecting the GKA semi-empirical rules. For $\Delta_{CT} << 0$, $J$ is no longer quartic, but quadratic power of $t^{(dp)}$ in 180$^{\circ}$ cases and is not a function of $t^{(dp)}$ but proportional to $j^{(p)}_H$ in 90$^{\circ}$ case. Thus 180$^{\circ}$ gives weak AFM while 90$^{\circ}$ gives strong FM in this situation, totally reversing the GKA semi-empirical rules.

Such a paradigm shift stems from the fact that $|d^{2}\underline{L}>$ has larger weight than $|d^{1}L>$ in NCTI as illustrated by {\color{blue}Eq. \ref{eq:mbs}}. For 180$^{\circ}$ with $\Delta_{CT}>> 0$, the zero-th order state for $S$ = 0 and $S$ = 1 have the same energy, as shown in {\color{blue}Fig. \ref{fig:cluster}(c)-(d)}. Since hopping process marked by \textcircled{1} are allowed for both $S$ = 0 and $S$ = 1, they have the same energy at this level of perturbation (see {\color{blue}Fig. \ref{fig:cluster}(g)-(h)}). The energy difference between $S$ = 0 and $S$ = 1 comes from the second-order perturbation where hopping process \textcircled{2} is allowed in $S$ = 0 while forbidden in $S$ = 1 (see {\color{blue}Fig. \ref{fig:cluster}(c)-(d)}). In this way, $S$ = 0 has lower energy than $S$ = 1 with $J$ positive and about $(t^{dp})^{4}$. The situation is almost the same for 90$^{\circ}$ with $\Delta_{CT} >>0$, but when \textcircled{1} and \textcircled{2} (see {\color{blue}Fig. \ref{fig:cluster}(e)-(f)}) are finished, the state in {\color{blue}Fig. \ref{fig:cluster}(i)} has higher energy than {\color{blue}Fig. \ref{fig:cluster}(j)} by the amount of $j^{(p)}_H$, which gives a weak negative $J$. 

If $\Delta_{CT} << 0$, the zero-th order state for both geometry will be changed. For 180$^{\circ}$, it is {\color{blue}Fig. \ref{fig:cluster}(g)-(h)} for $S$ = 0 and $S$ = 1 sector, which is just the state with \textcircled{1} finished in {\color{blue}Fig. \ref{fig:cluster}(c)-(d)} (here we insist $\Delta_{CT} > - u_p$ to void empty p orbital in the zero-th order state). Again, at the zero-th order, $S$ = 0 and $S$ = 1 have the same energy, but at the first order we can see the energy difference as hopping process \textcircled{2} is allowed in {\color{blue}Fig. \ref{fig:cluster}(g)} but forbidden in {\color{blue}Fig. \ref{fig:cluster}(h)}. That's the reason why $J$ is $\sim$ $t^{2}_{dp}$. While for 90$^{\circ}$ case, there is already an energy difference at the zero-th order as seen from {\color{blue}Fig. \ref{fig:cluster}(i)-(j)}, which gives a negative $J$ $\sim$ $j^{(p)}_H$ as listed in {\color{blue}Tab. \ref{tab:ana-J}}. 

From the above discussion, it is clear that negative CTI states lead to strong FM in the system. To further see the effect of band structure introduced by $t^{(d)}$, $t^{p}$, next we are going to study the phase diagram on two-dimensional (2D) bipartite square lattice on the mean-field level \cite{Claveau2014}.

\begin{figure*}
\centering
\includegraphics[width=16cm]{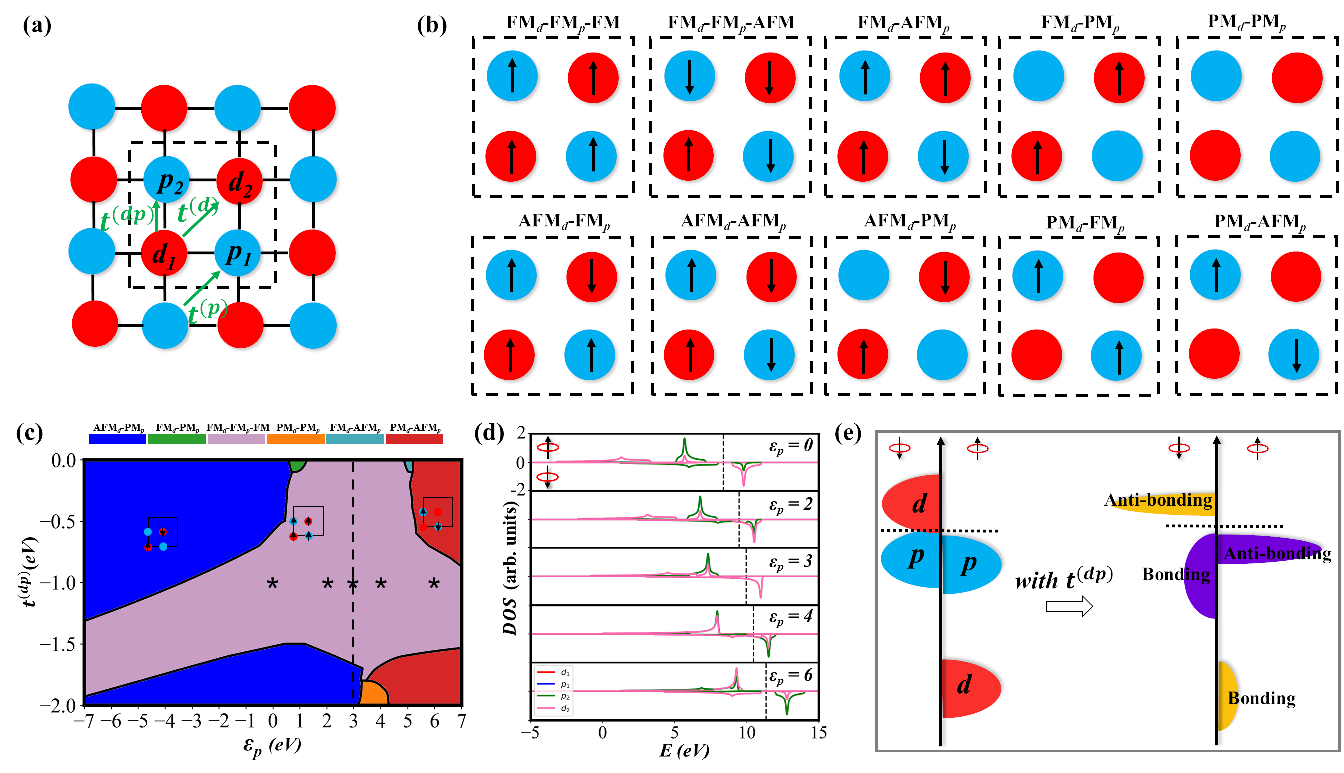}
\caption{(a) Bipartite square lattice. The blue/red filled circle represents $p$/$d$ orbital. The dashed square represents a $\sqrt{2} \times \sqrt{2}$ supercell. (b) Initial magnetic phases used in the mean-field calculation. (c) Phase diagram, here $\varepsilon_d$ = 0, $u_d$ = 9 eV and $u_p$ = 6 eV. (d) Partial DOS of points marked by stars in (c). (e) Schematic representation of hybridization gap.}
\label{fig:mf}
\end{figure*}

\subsection{C: Mean-field study of bipartite square lattice}
The bipartite square lattice is shown in {\color{blue}Fig. \ref{fig:mf}(a)}. To study both FM and ($\pi$, $\pi$)-AFM state, here a $\sqrt{2} \times \sqrt{2}$ supercell is applied. Here we fix $\varepsilon_d$ = 0 and vary $\varepsilon_p$. For hopping terms, $t^{(d)}$ and $t^{(d)}$ are fixed to be -0.5 eV (i.e. W$_d$ = W$_p$ = 4 eV) and $t^{(dp)}$ is varied. For the interacting parameter, we fix $u_d$ = 9 eV and $u_p$ = 6 eV so $u_d/|t^{(d)}|$ and $u_p/|t^{(p)}|$ is 18 and 12. With each $d$/$p$ orbital contributes 1/2 electrons, the total electron filling ($N_e$) in the supercell is 6.0. A 100 $\times$ 100 k-point grid is applied for integration and the temperature is set to be 30 K. During the mean-field simulation, 10 initial magnetic phases are applied as shown in {\color{blue}Fig. \ref{fig:mf}(b)}. The label FM$_d$-FM$_p$-FM means that both $d$ and $p$ sublattices form FM order and the whole magnetic order is FM. The threshold for charge self-consistency is 10$^{-5}$. 

The calculated phase diagram is shown in {\color{blue}Fig. \ref{fig:mf}(c)}. When $\varepsilon_p$ is far below $\varepsilon_d$, it is $d$ orbital that is active which forms AFM order as $d$ is half filled. Therefore the ground state is AFM$_d$-PM$_p$. On the contrary, when $\varepsilon_p$ is much larger than $\varepsilon_d$, it is $p$ orbital that is active which gives a PM$_d$-AFM$_p$ ground state (as $p$ is also treated as correlated orbital here). These two extreme cases are in accordance with the perturbative treatment of cluster model. What's intriguing is that FM$_d$-FM$_p$-FM straddles with intermediate $\varepsilon_p$ and such a phase is roughly symmetric with respect to $\varepsilon_p$ = 3 eV ( or $\Delta_{CT}$ = 0 eV) as shown in {\color{blue}Fig. \ref{fig:mf}(c)}. Such a phenomena can be understood as follows: with $\Delta_{CT}$ close to zero, either $d$ or $p$ orbital is away from fully occupied (especially all long-range magnetic orders are metallic when $\Delta_{CT}$ = 0 eV) which gives the possibility of  FM order in both $d$ and $p$ orbitals. This is further confirmed by the fact that with large $-t^{(dp)}$, FM$_d$-FM$_p$-FM can sustain in region with large |$\varepsilon_p$|. Although FM$_d$-FM$_p$-FM at $\varepsilon_p$ = 3 eV is metallic, insulating phase can be find away from $\varepsilon_p$ = 3 eV. {\color{blue}Fig. \ref{fig:mf}(b)} displays the spin resolved partial density of state (DOS) of FM$_d$-FM$_p$-FM phase for different $\varepsilon_p$ (marked as stars in {\color{blue}Fig. \ref{fig:mf}(c)} where $t^{(dp)}$ is fixed at -1.0 eV). When $\varepsilon_p$ is close to 3 eV, the system remains metallic. Nevertheless, when $\varepsilon_p$ is 0 or 6 eV, the system is insulating. The mechanism for gap opening at $\varepsilon_p$ = 0 eV is orbital hybridization, as illustrated in {\color{blue}Fig. \ref{fig:mf}(e)}. With $\varepsilon_p$ = 0 eV, $\varepsilon_{d,\downarrow}$ - $\varepsilon_{p,\downarrow}$ $\sim$ $(W_d + W_p)/2$, so a moderate $t^{(dp)}$ is able to open a large gap. However, if $\varepsilon_{d,\downarrow}$ - $\varepsilon_{p,\downarrow} >> (W_d + W_p)/2$, then the system will become a CTI which prefers AFM$_d$-PM$_p$ order. Meanwhile, if if $\varepsilon_{d,\downarrow}$ - $\varepsilon_{p,\downarrow} << (W_d + W_p)/2$, then huge $t^{(dp)}$ is required to open a gap which will drive the system to other magnetic orders.

In summary, with $\Delta_{CT}$ negative or close to zero, the exotic FM$_d$-FM$_p$-FM emerges and a moderate $t^{(dp)}$ can open a hybridization gap under the condition that $\varepsilon_{d,\downarrow}$ - $\varepsilon_{p,\downarrow}$ $\sim$ $(W_d + W_p)/2$.


\subsection{D: Ferromagnetic negative charge transfer insulator}
\begin{figure}
\centering
\includegraphics[width=8cm]{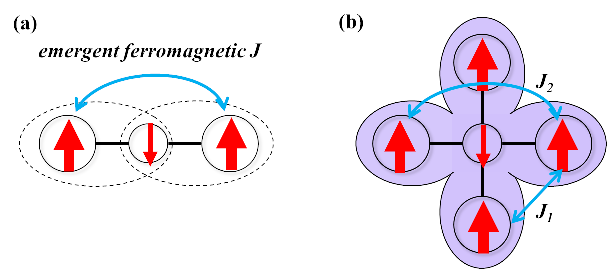}
\caption{(a) Schematic representation of emergent FM exchange coupling mediated by hole on $L$ $p$ orbital. The black dashed line represent strong AFM direct exchange coupling between M and $L$. (b) Magnetic molecular orbital with wave function extending over next-nearest neighbors, both $J_1$ and $J_2$ are FM. The large (small) circle represent M $d$ ($L$ $p$) orbital.}
\label{fig:emFM}
\end{figure}

Here we would like to call the ferromagnetic insulator like {\color{blue}Fig. \ref{fig:mf}(e)} "ferromagnetic negative charge-transfer insulator". Compared with FM CTI or MHI, here the origin of gap is due to $d$-$p$ orbital hybridization \cite{Nimkar1993}, rather than electron-electron correlations. Due to $d$-$p$ hybridization, the obtaining bonding and anti-bonding orbitals in spin down channel of {\color{blue}Fig. \ref{fig:mf}(e)} are no longer localized $d$/$p$ orbitals, but magnetic molecular orbitals (MMOs) which are linear combination of local $d$ and $p$ orbitals. These MMOs are non-local with large orbital spreadings and can reproduce the origin idea of Khomskii and Sawatzky \cite{Khomskii1997}. Taken a bipartite square lattice formed by M and $L$ (shown in {\color{blue}Fig. \ref{fig:emFM}(b)}) as example, both the nearest (NN) and next-nearest neighbor (NNN) superexchange interaction ($J_1$ and $J_2$) are emergent FM according to Khomskii and Sawatzky \cite{Khomskii1997}. In FNCTI here, we are treating a MMO whose orbital spreading has a characteristic length covering both NN and NNN. Within this characteristic length, all exchange coupling should be FM and thus both $J_1$ and $J_2$ are FM, in accordance with argument from Khomskii and Sawatzky \cite{Khomskii1997}.

To find FNCTI in real transition metal compounds, a multi-band extension is required to take multiple $d$ filling into consideration. At first glance, it is reasonable to assume that all $d$ ($p$) orbitals have the same $\varepsilon_d$ ($\varepsilon_d$) as the orbital splitting induced by point charges are generally small, therefore $j^{(d)}_H$ ($j^{(p)}_H$) will align spin polarization on $d$ ($p$) shell to high spin configuration, making the picture in {\color{blue}Fig. \ref{fig:mf}(e)} still valid except that Fermi level can be higher. This makes two major differences. Firstly, from {\color{blue}Fig. \ref{fig:mf}(e)}, it is clear that not all $d$ filling can give a gap order so we have the idea of "ideal $d$ filling": only at this $d$ filling that the system exhibits a gap. When the $d$ filling departs from "ideal $d$ filling", extra electron/hole then enters anti-bonding/bonding orbital of spin down channel, which destabilize the FM order. In this sense, the FM and gap order are synergistic in FNCTI. Secondly, we need to consider another case when parts of $d$ orbitals are empty. A simple calculation show that when $\Delta_{CT}$ is close to zero or negative, FM$_d$-FM$_p$-AFM will be the preferred ground state rather than FM$_d$-FM$_p$-FM (see {\color{blue}Appendix C}). It is noted that if the local spin of M is 1/2, the coupling between $d$ and $p$ can lead to bounded Zhang-Rice singlets \cite{Zhang1988} which makes the system paramagnetic (PM). Therefore it is better for local spin of M to be in the classical limit. In the following, we will consider this situation only. 

Strictly speaking, a semiconductor with FM$_d$-FM$_p$-AFM order should be coined as "ferrimagnetism" rather than ferromagnetism. But such a ferrinagnetism is inborn in the mechanism of FNCTI, which is slightly different from normal ferrimagnetism where the underlying local spins come from different transition metal cations. For this reason, we still use "ferromagnetism" in FNCTI.

\section{III. Material realization}
\subsection{A: Material candidate}
In this section, we look for material realization of FNCTI. In principle, FNCTI can be realized in any lattice structure. Here we choose PbO-type M$L$ monolayer as an example. Such a binary lattice structure resembles the bipartite square lattice: the M atoms form a square lattice with $L$ positioned in the middle of each square, alternatively above or below the paper plane. What's more, it is shared by ${\rm ThCr_2Si_2}$-\cite{ThCr2Si21, ThCr2Si22, ThCr2Si23, ThCr2Si24} and ZrCuSiAs-family materials \cite{ZrCuSiAs1, ZrCuSiAs2, ZrCuSiAs3, ZrCuSiAs4} and has been widely studied due to the raising of Fe-based superconductors \cite{Kamihara2008, Hsu2008, Wang2012}. 

For the M-$L$ combinations, M = V, Cr, Mn and $L$ = P, As, Sb are selected. In transition metal oxides, it is well known that $\Delta_{CT}$ systematically decreases with increasing atomic number or increasing formal valence of the metallic ions \cite{Bocquet-1992}. For 3$d$ metal cations with high-oxidation state (Cr$^{4+}$, Co$^{3+}$, Ni$^{3+}$, Cu$^{3+}$), $\Delta_{CT}$ can be very small or even negative. Therefore we choose $L$ to be pnictogen family, which gives M a high oxidation ${\rm M^{3+}}$. Nitrogen is ignored on purpose for its strong ionicity and weak covalent bonding with M.
As for the metal ions, V, Cr, Mn are chosen to tune $d$ filling (n)  2 $\sim$ 4. Here we focus on 3${d}$ transition metals for two reasons: firstly, they are lighter than their 4${d}$ and 5${d}$ cousins, the relativistic effect will be much weaker and won't drive the FM phase to other phases like quantum spin liquid \cite{SL2014}. Secondly, 3$d$ orbitals has large $j^{(d)}_H$ than 4${d}$ and 5${d}$, therefore high spin state is favored when multiple spin configurations are possible \cite{Cao2018}.

\begin{figure*}
\centering
\includegraphics[width=16cm]{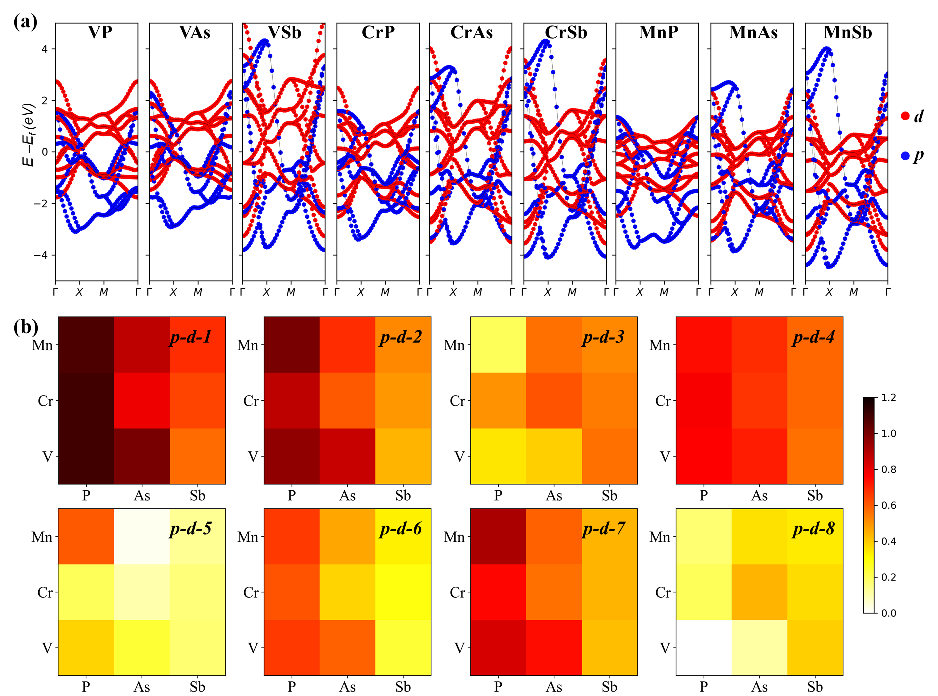}
\caption{(a) Band structure of PM M$L$ monolayer without $t^{(dp)}$. The high symmetric k-path is $\rm \Gamma$-X-M-$\rm \Gamma$: (0.0, 0.0)-(0.5, 0.0)-(0.5, 0.5)-(0.0, 0.0). (b) Strength of $t^{(dp)}$ in different $p-d$ hopping channel according to {\color{blue}Tab. \ref{Tab:tdp}}. To avoid  phase dependence of $t^{(dp)}$, the phase is chosen so that all $t^{(dp)}$ is positive. The unit is eV.}
\label{fig:spp}
\end{figure*}

\subsection{B: Paramagnetic phase}
We first turn to the PM phase. The fitted single-particle parameter  is shown in {\color{blue}Fig. \ref{fig:spp}} (see {\color{blue}Appendix D} for model parameters fitting). {\color{blue}Fig. \ref{fig:spp}(a)} shows the band structure without $t^{(dp)}$. Due to the high oxidation state of M, $p$ orbitals are entangled with $d$ orbitals, making the $\varepsilon_{dp}$ close to 0 or even negative. When $L$ goes from P to Sb, both $W_p$ and $W_d$ get much wider (from 4 to 8 eV). 
{\color{blue}Fig. \ref{fig:spp}}(b) displays the strength of different $p-d$ hybridization channels (see {\color{blue}Fig. \ref{Fig:PBE}(c)} and {\color{blue}Tab. \ref{Tab:tdp}} for the definition). The $p-d-5$ is the strongest in the buckling-free case such as in CuO$_2$ plane, but now it becomes almost the smallest (only larger than $p-d-8$). Thus such buckling can not be regarded as little geometrical deviation here. The strongest $p-d$ hybridization now becomes the $p-d-1$ channel, which can be as large as 1.1 eV in VP and CrP, such a large ${\sigma}$-type hopping is quite astonishing. The second largest comes from the $p-d-2$, which is 1.0 eV in VP and MnP. Such large multiple $p-d$ hybridizations are essential to open a hybridization gap considering the large $W_p$ and $W_d$.
Seen from {\color{blue}Fig. \ref{fig:spp}(b)}, as $L$ goes from P to Sb, nearly all $p-d$ hybridizations become smaller. Considering the fact that $W_p$ and $W_d$ get wider from P to Sb, it will be harder for MSb to open a gap than MP and MAs in the FM phase.

\begin{table}
\setlength{\belowcaptionskip}{10pt}
\centering
\caption{Interaction parameters obtained from cRPA calculation. The unit is eV}
\begin{tabular}{p{10mm}<{\centering}p{8mm}<{\centering}p{8mm}<{\centering}p{8mm}<{\centering}p{8mm}<{\centering}p{8mm}<{\centering}p{8mm}<{\centering}p{8mm}}
\toprule
System  &    $u_d$  & $u'_d$ & $j^{(d)}_H$ & $u_p$  & $u'_p$ & $j^{(p)}_H$ & $u_{dp}$ \\
\midrule
VP     & 9.34 & 8.24 & 0.55 & 6.60 & 5.70 & 0.40 & 3.90  \\
VAs    & 9.31 & 8.19 & 0.56 & 6.04 & 5.21 & 0.38 & 3.65  \\
VSb    & 7.77 & 6.72 & 0.53 & 5.22 & 4.54 & 0.40 & 2.97  \\
CrP    & 9.60 & 8.39 & 0.61 & 6.80 & 5.86 & 0.45 & 3.95  \\
CrAs   & 8.89 & 7.67 & 0.61 & 6.08 & 5.25 & 0.46 & 3.37  \\
CrSb   & 7.16 & 5.90 & 0.64 & 4.41 & 3.64 & 0.41 & 1.88  \\
MnP    & 8.80 & 7.55 & 0.63 & 6.38 & 5.40 & 0.42 & 3.74  \\
MnAs   & 9.39 & 8.07 & 0.66 & 6.34 & 5.47 & 0.47 & 3.66  \\
MnSb   & 6.52 & 5.18 & 0.68 & 4.04 & 3.28 & 0.41 & 1.62  \\
\bottomrule
\end{tabular}
\label{tab:cRPA}
\end{table}

{\color{blue}Tab. \ref{tab:cRPA}} lists all the interaction parameters for the nine M$L$ monolayers. The $u_d$ is the leading energy scale which is around 9 eV. The $u_p$ term, the Hubbard interaction on $p$ shell, is smaller but can be as large as 6 eV. Therefore the electron-electron correlations on $p$ are large enough to form a sub-magnetic order. With $L$ becoming heavier, $u_d$ and $u_p$ undergo large decreases. For example, when we go from MnAs to MnSb, the reduction of $u_d$ ($u_p$) is up to 2.8 eV (2.3 eV). On the contrary, the Hund's coupling $j^{(d)}_H$ and $j^{(p)}_H$ are less influenced by the elemental differences and remain values proximate to 0.60 and 0.40 eV. Another noticeable fact is the large inter-shell Coulomb interaction $u_{dp}$, such a value is much larger than that in iron-based superconductors \cite{Roekeghem2016} and comparable with that expected in cuprates \cite{Hansmann2014}. 

\begin{figure*}
\includegraphics[width=16cm]{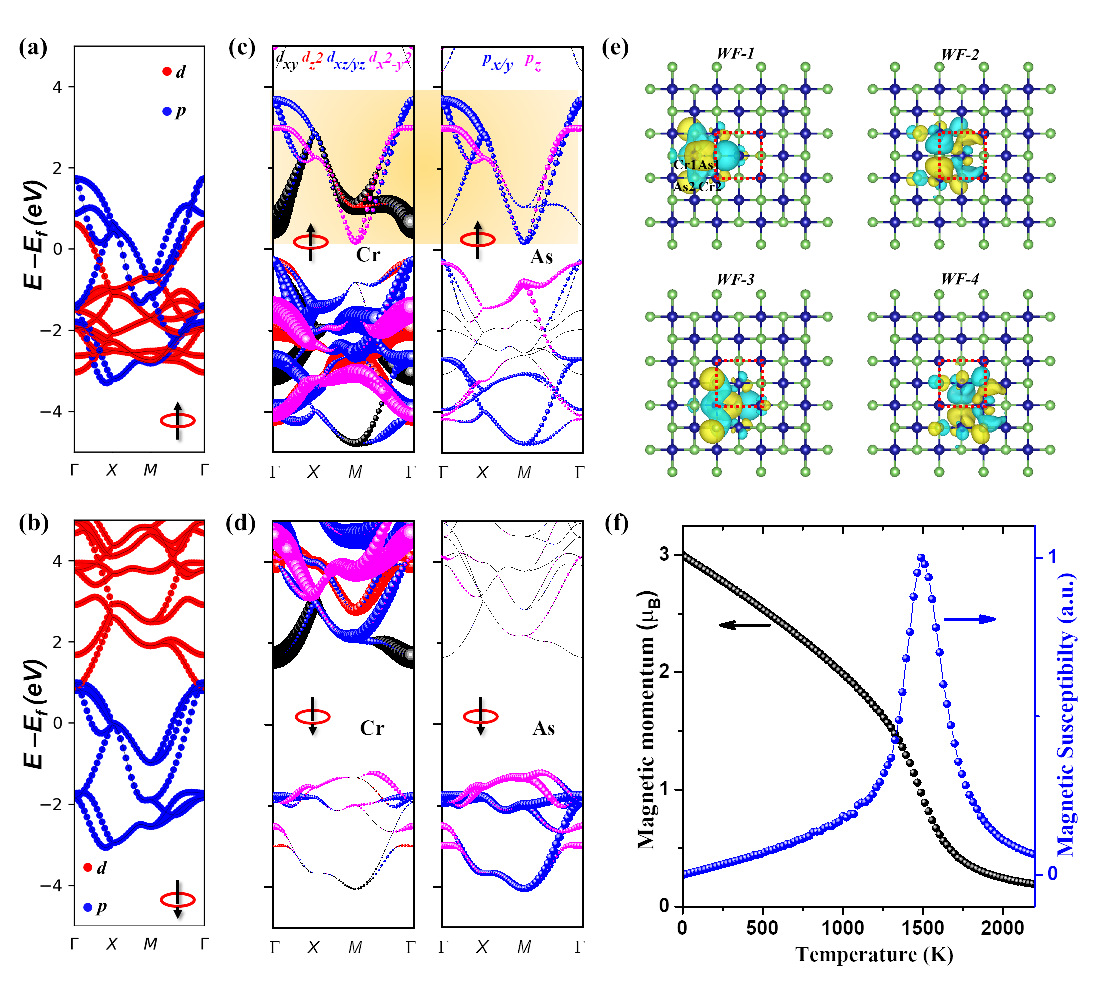}
\caption{(a)-(b) Band structure of FM CrAs monolayer without $t^{(dp)}$ for spin up/down channel. (c)-(d) Orbital-resolved band structure for spin up/down channel. The left/right panel is for Cr/As respectively. The anti-bonding bands are marked by shaded yellow region in (c). In (c)-(d), different colors are used to stand for different atomic orbitals. (e) The four maximally localized Wannier functions for the anti-bonding orbitals for up spin. Red dashed rectangle represents unit cell. (f) Evolution of effective magnetic moment (black) and magnetic susceptibility (blue) with respect to temperature.}
\label{fig:fmCrAs}
\end{figure*}

\subsection{C: Ferromagnetic phase}
\subsubsection{1. n = 3}
We first consider n = 3 and take CrAs as an example. Without $p-d$ hybridization, both spins are metal as shown in {\color{blue}Fig. \ref{fig:fmCrAs}(a)-(b)}. Compared with PM, $\varepsilon_{dp}$ experiences huge changes: spin up $p$ is still in highly entanglement with $d$ orbitals with several $p$ bands higher than $d$ bands while spin down $p$ is well separated from $d$ band. 
With $p-d$ hybridization turning on, an indirect (direct) gap $\sim$ 0.35 eV (2.82 eV) opens in the spin up (down) channel as shown in {\color{blue}Fig. \ref{fig:fmCrAs}(c)-(d)}, which makes FM CrAs monolayer a semiconductor. The origin of gap in different spin channels are different, as can be seen in the orbital projected band structure in {\color{blue}Fig. \ref{fig:fmCrAs}(c)-(d)}. For spin up channel, both $p$ and $d$ orbitals have competing weights on the anti-bonding orbitals (mainly formed by $p_x, p_y, d_{xz}, d_{yz}, d_{xy}$ as marked by light yellow region) and several bonding orbitals, therefore the origin of gap here is orbital hybridization. Due to the large $t^{(dp)}$ between $p_x$ ($p_y$) and $d_{xz}$ ($d_{yz}$) in channel $p-d-1$ and between $p_x + p_y$ and $d_{xy}$ in channel $p-d-2$, we can see a global gap in the spin up channel here. As for spin down channel, the gap is mainly assigned to the large $\varepsilon_{dp}$, as the anti-bonding (bonding) orbitals are dominated by the Cr 3$d$ (As 3$p$) orbitals. 

To see the magnetic molecular orbitals in the spin up channel, here we downfold the 4 anti-bonding magnetic Wannier functions (MWFs) and the results are plotted in {\color{blue}Fig. \ref{fig:fmCrAs}(e)}. We find that the these MWFs consist of small MX clusters, rather than individual M or $L$ atoms. To be specific, WF-1 is formed by As1 $p_x$, Cr2 $d_{xy}$ and Cr1 $d_{xz}$ while WF-2 is constituted by As1 $p_y$, Cr2 $d_{yz}$ and Cr1 $d_{xy}$, the other two WFs are equivalent to WF-1 and WF-2. Such a linear combination of orbitals is in accordance with the orbital composition of these bands depicted in {\color{blue}Fig. \ref{fig:fmCrAs}(c)}. Two consequences come along: at first, $p$ orbitals are populated with non-negligible holes which gives a local magnetic moment -0.42 $\mu_B$. Secondly, WFs are highly non-local and even cover the NNN Cr dimmers as plotted in {\color{blue}Fig. \ref{fig:emFM}(b)}. As discussed in {\color{blue}Section-II}, both $J_1$ and $J_2$ should be FM. 

\begin{figure*}
\includegraphics[width=16cm]{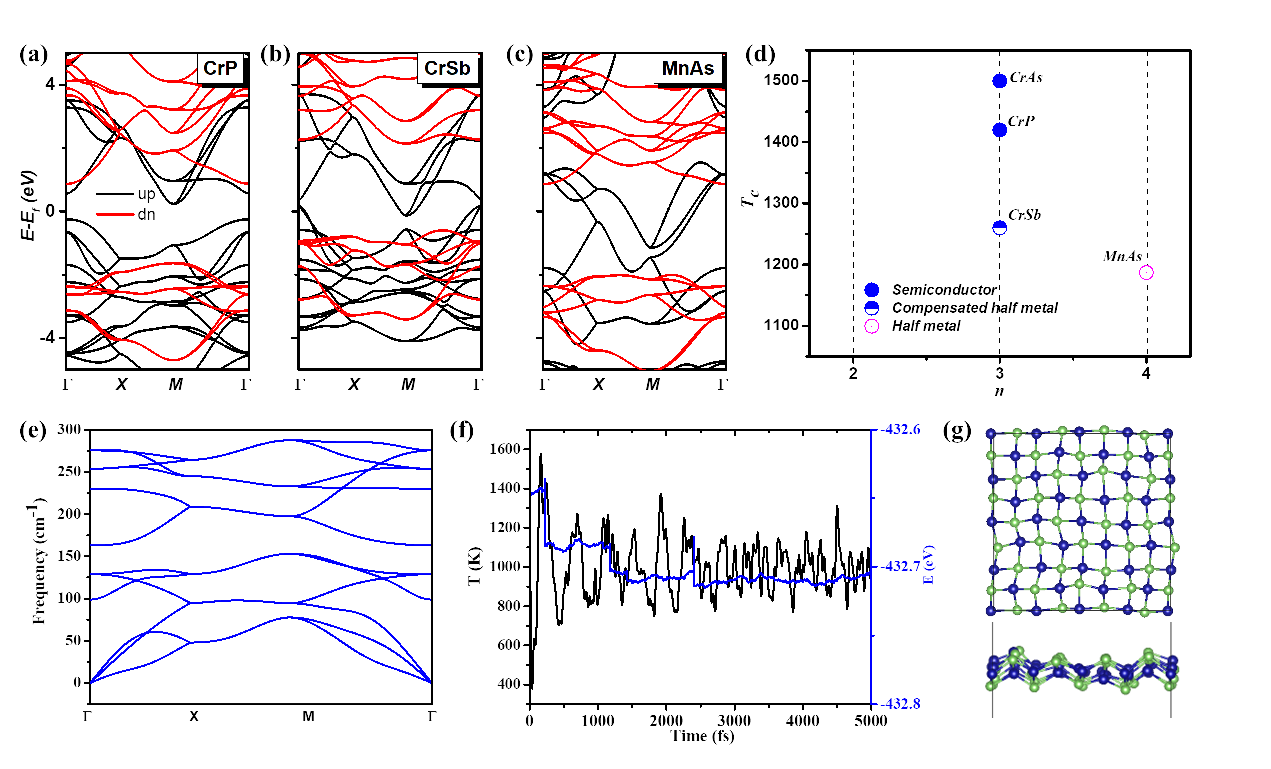}
\caption{(a)-(c) Band structure of FM CrP, CrAs and MnAs monolayer. (d) A schematic phase diagram with respect to n. (e) Phonon spectrum of FM CrAs monolayer. (f) AIMD simulation of FM CrAs monolayer at 1000 K. (g) Top and side view of snapshot taken at 5 ps in AIMD simulation.}
\label{fig:fmother}
\end{figure*}

According to energy mapping method (see {\color{blue}Appendix E}), the calculated $J_1$ and $J_2$ are -56.8 and -39.7 meV, both are ferromagnetic. In FM MHI and CTI, one expects the $J_{ex}$ to be rather short-range, the real surprise here is the magnitude of $J_2$ over $J_1$. The ratio $J_2$:$J_1$ $\sim$ $1/\sqrt{2}$:1 strongly implies linear-law scaling here. As both $J_1$ and $J_2$ are ferromagnetic, the magnetic ground state in such a square lattice is FM, in accordance with first-principles calculations. To determine $T_c$, classical Monte Carlo (MC) simulations are performed for a ${32 \times 32 \times 1}$ supercell based on Heisenberg Hamiltonian with $J_1$ and $J_2$ \cite{PASP}. During FM-PM phase transition, magnetic susceptibility is calculated after the system reaches equilibrium at a given temperature, then $T_c$ corresponds to the position of peak in magnetic susceptibility plot as shown in {\color{blue}Fig. \ref{fig:fmCrAs}(f)}. As anticipated, an ultra-high $T_c$ $\sim$ 1500 K is obtained. For 2D magnets, strong magnetic anisotropy energy (MAE) is needed to break the Hohenberg-Merin-Wagner theorem \cite{Hohenberg, Mermin}. Since the highest occupied and lowest unoccupied orbital are mainly contributed by Cr ${d_{xz/yz}}$ and $d_{xy}$ in spin up (see {\color{blue}Fig. \ref{fig:fmCrAs}(c)}), the orbital angular momentum difference ($||\Delta L_{z}||$) between these orbitals is 1, the prefer spin orientations of magnetic ions should be perpendicular to out-of-plane direction \cite{MAE1, MAE2}. To verify this point, spin-orbit coupling (SOC) is taken into account in calculating the relative energies along (001) and (100) direction. It is (100) that is the easy axis, in accordance with orbital composition analysis. The calculated MAE is $\sim$ 0.37 meV/Cr, comparable to that in ${\rm CrI_3}$ \cite{MAE3}.

Now we count the electron filling. For the 18 electrons from Cr $d$ and As $p$ orbitals, 6 of them occupy the spin down bands and the left 12 electrons occupy the spin up bands. Seen from {\color{blue}Fig. \ref{fig:fmCrAs}(c)}, these 12 electrons just fill bands up to the gap which makes CrAs monolayer a semiconductor. In this sense, n = 3 is "ideal $d$ filling" for such a lattice structure. Along this logic, CrP and CrSb monolayer should also belong to FNCTI. {\color{blue}Fig. \ref{fig:fmother}(a)-(b)} shows the band structure of FM CrP and CrSb monolayer. CrP monolayer resembles CrAs monolayer very much, which is also a small gap semiconductor with similar $J_1$ and $J_2$ as displayed in {\color{blue}Tab. \ref{tab:JC}}. As for CrSb monolayer, because of larger $W_p$ ($W_d$) and smaller $t^{(dp)}$, there is no global gap in {\color{blue}Fig. \ref{fig:fmother}(b)}, which makes CrSb monolayer a compensated half-metal. The calculated $J_1$ and $J_2$ (see {\color{blue}Tab. \ref{tab:JC}}) of CrSb monolayer are smaller than CrAs and CrP monolayer which gives a slightly lower $T_c$ $\sim$ 1260 K. 

\subsubsection{2. other filling}
VL and MnL can be regarded as one hole and electron doped CrL per unit cell. Generally speaking, the extra charge doping will mediate  FM exchange interactions between localized electrons, thus enhancing FM in the system. Nevertheless, from {\color{blue}Fig. \ref{fig:mf}(e)}, the doping of holes (electrons) will decrease (increase) the occupation of bonding (anti-bonding) states, hence destabilizing the FM ground state. The compete between these two factors may drive FM-AFM transition. To see the robustness of FM in FNCTI, here we go on studying VL and MnL monolayers. For VL, no magnetic orders are found, which maybe due to the small local magnetic moment on V. For MnL, FM solutions are found for MnAs while MnP and MnSb prefers AFM ground state \cite{Wang2019}. The band structure of FM MnAs monolayer is depicted in {\color{blue}Fig. \ref{fig:fmother}(c)} which is a half-metal. {\color{blue}Tab. \ref{tab:JC}} lists $J_1$ and $J_2$ value of MnAs, it is clear that $J_1$ is much smaller than that in CrL, suggesting the destabilization effect is much stronger than itinerant enhancement. Such a phenomena can be taken as an indicator for FNCTI in experiment. With large $J_2$ and larger magnetic moments, the $T_c$ of MnAs monolayer is still over 1100 K.

{\color{blue}Fig. \ref{fig:fmother}(d)} summarizes the results for all the FM monolayers. For the "ideal $d$ filling" n = 3, FNCTI is found in CrAs and CrP monolayer, a nearby phase of FNCTI is FM compensated half-metal, as found in CrSb monolayer. Deviating from ideal filling, no FM is obtained in n = 2 while FM half-metal phase is possible for n = 4, like in MnAs monolayer.  

\begin{table}
\setlength{\belowcaptionskip}{10pt}
\centering
\caption{ Exchange coupling strength and estimated $T_c$ for different FM monolayers}
\begin{tabular}{p{28mm}<{\centering}p{14mm}<{\centering}p{16mm}<{\centering}p{18mm}<{\centering}}
\toprule
System  &   $J_1$ (meV)    & $J_2$ (meV)  & $T_c$ (K) \\
\midrule
CrP        & -55.3   & -36.2  & 1420  \\
CrAs       & -56.8   & -39.7  & 1500  \\
CrSb       & -51.7   & -25.8  & 1260  \\
MnAs       & -11.7   & -32.5  & 1187  \\
\bottomrule
\end{tabular}
\label{tab:JC}
\end{table}

\subsubsection{3. Stability of CrAs monolayer}
The existence of bulk CrAs in nature \cite{bulk1, bulk2} suggests the 1:1 stoichiometric ratio in CrAs monolayer is charge feasible. The stability of CrAs monolayer is confirmed by both phonon spectrum and \textit{ab-initio} molecular dynamics (AIMD) ({\color{blue}Fig. \ref{fig:fmother}(e)-(g)}), in accordance with previous report \cite{Ma2020}. The stability indicates an experimental preparation of CrAs monolayer is possible. With layered BaCr$_2$As$_2$ \cite{ThCr2Si23} and LaCrAsO \cite{ZrCuSiAs4} synthesized in experiments, CrAs monolayer can be obtained by either etching \cite{MXene} or electrochemical reactions \cite{eleexo} from these layered materials. Another synthetic method is molecular-beam epitaxy and ${\rm BaZrO_3}$ (001) (or MgO (001)) is a perfect substrate with 1:1 lattice match. Such a strategy has shown success in CoSb monolayer preparation \cite{mbe1, mbe2}. 

\section{V. Discussion and Conclusion}
Up to here, we illustrate the concept of FNCTI and suggest the material realization in CrAs monolayer. FNCTI are different from FM CTI and MHI from the following aspects:  
\textit{(1) Origin of gap}. In FM CTI and MHI, the gap is originated from electron-electron correlation and is always large. However in FNCTI, the gap comes from $d$-$p$ hybridization, which is commonly small. Since strong orbital hybridization reflects the co-valency nature of the system, FNCTI thus lies at a special point where kinetic and interaction energy make peace. In this regard, HSE06 functional is a good functional to study FNCTI from perspective of first-principles calculation (See {\color{blue}Appendix A} for a detailed discussion).
\textit{(2) Scaling behavior of $J$}. In FM CTI and MHI, the building motifs of magnetic interaction are local atomic orbitals with nearly fully filled $p$ orbitals, so $J$ decays quickly with respect to distance between magnetic pairs. However in FNCTI, the building motifs are MMOs, theirs large orbital extension gives slowly decaying $J$ within MMOs and sudden decrease beyond MMOs (see {\color{blue}Appendix F}). 
\textit{(3) Response to electron/hole doping}. In FM CTI and MHI, electron or hole doping tends to bring itinerant FM into the system and thus enhancing FM. While in FNCTI, extra electron or hole will destabilize the pristine FM phase. This may explain why LaCrAsO occupies AFM order \cite{ZrCuSiAs4}. Replacing La by Sr, part of Cr$^{2+}$ will become Cr$^{3+}$ and a AFM-FM transition may be observed.

It is well known that $W_d$ and $W_p$ will becomes larger as the system grows from 2D to 3D, thus FNCTI is hard to be found in pure 3D materials. In 2D materials, because the band width and $d$-$p$ hybridization are easily tuned by external strain, a FM metal-to-insulator phase transition can be achieved via strain engineering. Such a phenomena would be hard to observe in FM CTI or FM MHI. What's more, as $p$ can be higher than $d$ orbitals in FNCTI, in view of band topology, this gives band inversion and non-trivial topology \cite{Devakul2022}. Therefore, FM negative CTI is also a good platform to study quantum anomalous Hall effect.

We believe our work also complete the family of negative CTI. PM negative CTI has been reported in NaCuO$_2$ \cite{NaCuO2-1991}, RNiO$_3$(R is a rare earth)\cite{RNiO3-2016}, and metallic FM negative CTI has been reported in CrO$_2$ \cite{CrO2-1998}, SrCoO$_3$ \cite{SrCoO3-2012}. Here CrAs monolayer represents an ideal example of (gapped) FNCTI. 

In conclusion, here a different type of FMSC is proposed: FNCTI. It breaks the GKA semi-empirical rules and the corresponding $T_c$ can be much higher than room temperature. Through first-principles calculation with hybrid functional, CrAs monolayer is mapped out to be a typical FNCTI, which has a band gap around 0.35 eV and a high $T_c$ about 1500 K. Due to its exotic physical properties, it is envisioned that FNCTI will arouse broad interest in condensed matter physics.

Z. L. thanks A. van Roekeghem and H. Jiang for helpful discussion. This work is supported by NSFC (No. 21688102, 12174356 and 22073086), 
National Key R\&D Program of China (No. 2017YFA0204904, 2016YFA0200604), Youth Innovation Promotion Association CAS (2019441) and the Start-up Funding from Westlake University. We thank Supercomputing Center at USTC for providing the computing resources.

\clearpage
\widetext
\appendix
\begin{appendices}

In this supplemental material, we provide more details of the calculation and results to support the discussion in the main text.
In sec. A, we make a brief introduction to the calculation details, especially we aim at explaining why the hybrid functional is a good choice for the study of FNCTI. 
In Sec. B, we derive the results in {\color{blue}Tab. \ref{tab:ana-J}}. This section includes three subsections.
In Sec. C, we extend the mean-field results to situations with empty $d$ orbitals. 
In Sec. D, we talk about the model parameter calculation, including the single-particle part and the interacting part.
In Sec. E, we discuss the energy mapping method where the exchange coupling strength $J$ are calculated.
In Sec. F, we focus on the scaling behavior of $J$, here a large supercell is applied and exchange coupling strengths up to $J_4$ are obtained.

\section{APPENDIX A: Calculation details and functional dependence}
\begin{figure*}
\includegraphics[width=16cm]{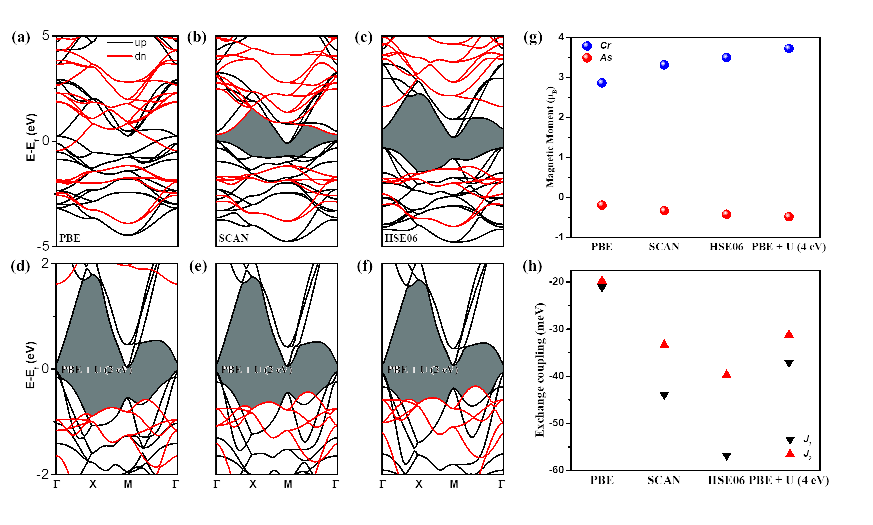}
\caption{Calculated band structure of FM CrAs monolayer by (a) PBE functional, (b) SCAN functional, (c) HSE06 functional and PBE + U method with (d) U = 3 eV, (e) U = 4 eV and (f) U = 5 eV. (g) Dependence of magnetic moments of Cr and As with respect to different functionals and methods. (h) Dependence of exchange coupling $J_1$ and $J_2$ with respect to different functionals and methods. }
\label{Fig:Functional}
\end{figure*}

Our first-principles calculations were performed on density functional theory implemented in the Vienna Ab initio Simulation Package (VASP) \cite{VASP}. For geometric optimization and electronic property calculations, a plane-wave cutoff 600 eV is used. The energy convergence criterion is $10^{-6}$ eV and the residual force is 0.01 eV/{\AA}. The Brillouin zone integration is carried out with $12 \times 12 \times 1$ k-point sampling for paramagnetic phase with PBE functional \cite{PBE}. The PBE functional is also applied in phonon spectrum calculation \cite{Togo2015} and AIMD simulation at 1000 K \cite{Martyna1992}.

To study the magnetic phase, a few remarks should be made here. As discussed in the main text both $p$ and $d$ orbitals are close to Fermi level, a suitable theory describing FNCHI should contain both interaction terms of both $p$ and $d$ orbitals. Especially, inter-shell interaction term is also needed to describe the charge correlations between $d$ and $p$ shells. Such a feature impose strong constraint on theoretical approach. In this sense, Heyd-Scuseria-Ernzerhof (HSE) functional \cite{HSE2003, HSE2004} should be better than functional such as PBE, SCAN \cite{SCAN2015, SCAN2016} or embedding method such as standard PBE + U \cite{LDAU1995, LDAU1997}. 

To see the performance of different functionals and methods, we use them to study the CrAs FM phase. The results are summarized in {\color{blue}Fig. \ref{Fig:Functional}}. We first consider the sequence of PBE, SCAN and HSE06 as they lie at the second, third and fourth rungs in the Jacob's ladder \cite{Perdew2005}. From {\color{blue}Fig. \ref{Fig:Functional}(a)}, PBE gives a itinerant FM phase, so the local magnetic moments on Cr and As are the smallest among these four functionals and methods. The SCAN functional is able to create a local gap for spin up channel, but is unable to create global gap ({\color{blue}Fig. \ref{Fig:Functional}(b)}, which makes CrAs a compensated FM metal. Due to the existence of local gap, the local magnetic moments on Cr and As are much larger than that of PBE. 
Finally, HSE06 corrects the gap to a global one $\sim$ 0.35 eV of spin up as shown in {\color{blue}Fig. \ref{Fig:Functional}(c)}. As the functional becomes more and more advanced as we go from PBE to SCAN to HSE06, the true magnetic properties are gradually approached.

Now we consider the PBE + U method, here we only add Hubbard U term on Cr 3$d$ shell and U = 3, 4 and 5 eV are applied to see the trend. Different U give a similar band structure as shown in {\color{blue}Fig. \ref{Fig:Functional}(d)-(f)}. As in HSE06 functional, a global gap is opened, but the band gap of spin up is only about 0.10 eV, much smaller than HSE06. The local magnetic moments of PBE + U method is much larger than SCAN and HSE06, indicating PBE + U tends to localize the electrons. At the same time, the insufficient description of kinetic energy of PBE + U gives a gap much smaller than than HSE06.

Combining the above two considerations together, it can be seen that HSE06 functional is the best functional (method) in describing both the valency and correlation character of FNCTI. In the main text, the magnetic properties, including band structure, exchange coupling strength estimation and magnetic isotropic energy are calculated using the HSE06 functional. For the $2 \times 2 \times 1$ supercell used in exchange coupling strength estimation, we use a $12 \times 12 \times 1$ k-point sampling. 

\newpage
\section{APPENDIX B: Derivation of {\color{blue}Tab. \ref{tab:ana-J}}}
In this Appendix, we derive the results in {\color{blue}Tab. \ref{tab:ana-J}}. In the following, the many-body state is written in the occupation representation. For 180$^{\circ}$ geometry shown in {\color{blue} Fig. \ref{fig:cluster}(a)}, the orbital order in the many-body state is $d_{\alpha}$, $p$ and $d_{\beta}$. For 90$^{\circ}$ geometry shown in {\color{blue} Fig. \ref{fig:cluster}(b)}, the orbital order in the many-body state is $d_{\alpha}$ , $p_x$, $p_y$ and $d_{\beta}$. For each orbital, $|0,0>, |\uparrow,0>, |0,\downarrow>, |\uparrow;\downarrow>$ represent empty state, singly occupied state with spin up/down and doubly occupied state, respectively. 

For a 2$\times$2 matrix as follows:
\begin{equation}
\left[
\begin{array}
{cc} 
H_{00}&T_{01}\\
T_{10}&H_{11}\\
\end{array}
\right]
\end{equation}
If we are interested in $H_{00}$ and the eigenvalues of $H_{11}$ are separated from $H_{00}$ with a large gap, we can integrate degrees of freedom of $H_{11}$ out by
\begin{equation}
H_{eff} = H_{00} + T_{01}(\epsilon - H_{11})^{-1}T_{10}
\label{Eq:per}
\end{equation}
where $\epsilon$ is the eigenvalue of original 2$\times$2 matrix. Approximating $\epsilon$ by eigenvalue of $H_{00}$, we obtain the effective matrix describing the degrees of freedom of $H_{00}$.

If there are one more high energy scale characterized by H$_22$, in other words we are treating a 2$\times$2 matrix as follows:
\begin{equation}
\left[
\begin{array}
{ccc} 
H_{00} &T_{01} &H_{02}\\
T_{10} &H_{11} &H_{12}\\
T_{20} &H_{21} &H_{22}\\
\end{array}
\right]
\end{equation}
By recursively applying {\color{blue}Eq. \ref{Eq:per}}, we have:
\begin{equation}
H_{eff} = H_{00} + T_{01}[\epsilon - (H_{11} + T_{12}(\epsilon - H_{22})^{-1}T_{21})]^{-1}T_{10}
\label{Eq:per2}
\end{equation}

The following approximation is often used to simplify the calculation of inverse of a matrix:
\begin{equation}
(A-B)^{-1} \sim A^{-1} + A^{-1} B A^{-1}
\label{Eq:per3}
\end{equation}
where the eigenvalues of A separate from that of B by a large gap.

\subsection{B-I: 180$^{\circ}$ geometry}
\subsubsection{B-I-1: $S$ = 0 sector}
There are 6 states in the $S$ = 0 sector: \\
$\{
\frac{1}{\sqrt{2}}(|\uparrow,0;\uparrow,\downarrow;0,\downarrow>-|0,\downarrow;\uparrow,\downarrow;\uparrow,0>), 
\frac{1}{\sqrt{2}}(|\uparrow,\downarrow;\uparrow,0;0,\downarrow>-|\uparrow,\downarrow;0,\downarrow;\uparrow,0>), 
\frac{1}{\sqrt{2}}(|\uparrow,0;0,\downarrow;\uparrow,\downarrow>-|0,\downarrow;\uparrow,0;\uparrow,\downarrow>), 
|\uparrow,\downarrow;\uparrow,\downarrow;0,0>, 
|0,0;\uparrow,\downarrow;\uparrow,\downarrow>, 
|\uparrow,\downarrow;0,0;\uparrow,\downarrow> 
\}$ \\
For example, the many-body state $\frac{1}{\sqrt{2}}(|\uparrow,0;\uparrow,\downarrow;0,\downarrow>-|0,\downarrow;\uparrow,\downarrow;\uparrow,0>)$ describes a single bond between $d_{\alpha}$ and $d_{\beta}$ and a doubly occupied $p$. 

For $\Delta_{CT}$ >> 0, the energy of $\frac{1}{\sqrt{2}}(|\uparrow,0;\uparrow,\downarrow;0,\downarrow>-|0,\downarrow;\uparrow,\downarrow;\uparrow,0>)$ is $2\epsilon_d + 2\epsilon_p + u_{p}$, which we will take as reference energy. Accordingly, the energy of $\frac{1}{\sqrt{2}}(|\uparrow,\downarrow;\uparrow,0;0,\downarrow>-|\uparrow,\downarrow;0,\downarrow;\uparrow,0>)$ is $\epsilon_d - \epsilon_p + u_d - u_p$, which is the charge transfer energy $\Delta_{CT}$ in this situation. At the same, the state $|\uparrow,\downarrow;\uparrow,\downarrow;0,0>$ has energy $u_d$. If $\Delta_{CT}$ > $u_d$, the system lies at MHI regime and the lowest excitation energy is $u_d$ rather than $\Delta_{CT}$. Here we are interested in CTI regime and assume $\Delta_{CT}$ < $u_d$.

The configuration interaction matrix in the basis $\{
\frac{1}{\sqrt{2}}(|\uparrow,0;\uparrow,\downarrow;0,\downarrow>-|0,\downarrow;\uparrow,\downarrow;\uparrow,0>), 
\frac{1}{\sqrt{2}}(|\uparrow,\downarrow;\uparrow,0;0,\downarrow>-|\uparrow,\downarrow;0,\downarrow;\uparrow,0>), 
\frac{1}{\sqrt{2}}(|\uparrow,0;0,\downarrow;\uparrow,\downarrow>-|0,\downarrow;\uparrow,0;\uparrow,\downarrow>), 
|\uparrow,\downarrow;\uparrow,\downarrow;0,0>, 
|0,0;\uparrow,\downarrow;\uparrow,\downarrow>, 
|\uparrow,\downarrow;0,0;\uparrow,\downarrow> 
\}$ is 
\begin{equation}
\left[
\begin{array}
{cccccc} 
0     &t^{(dp)} &t^{(dp)} &0               &0      &0              \\
t^{(dp)}&\Delta_{CT} &0      &-\sqrt{2}t^{(dp)} &0      &-\sqrt{2}t^{(dp)}\\
t^{(dp)}&0      &\Delta_{CT} &0      &-\sqrt{2}t^{(dp)} &-\sqrt{2}t^{(dp)}\\
0     &-\sqrt{2}t^{(dp)}&0      &u_d     &0      &0 \\
0     &0      &-\sqrt{2}t^{(dp)}&0       &u_d    &0 \\
0     &-\sqrt{2}t^{(dp)}&-\sqrt{2}t^{(dp)} &0      &0      &2\Delta_{CT}+u_p
\end{array}
\right]
\end{equation}

Here we are interested in $\epsilon \sim 0$ and take  
\begin{equation}
\begin{split}
H_{00} &= 0  \\
H_{11} &= 
\left[
\begin{array}
{cc} 
\Delta_{CT}    &0              \\
0         &\Delta_{CT}         \\
\end{array}
\right] \\
H_{22} &= 
\left[
\begin{array}
{ccc} 
u_d   &0       &0          \\
0     &u_d     &0          \\
0     &0       &2\Delta_{CT}+u_p\\
\end{array}
\right] \\
\end{split}
\end{equation}

By setting $\epsilon$ = 0, {\color{blue}Eq. \ref{Eq:per2}} becomes:
\begin{equation}
H_{eff} \approx H_{00} - T_{01}(H_{11} - T_{12}H^{-1}_{22}T_{21})^{-1}T_{10}
\end{equation}
By taking A = $H_{11}$ and B = $T_{12}H^{-1}_{22}T_{21}$ in {\color{blue}Eq. \ref{Eq:per3}}, we can further simplify the above equation as:
\begin{equation}
H_{eff} \approx H_{00} - T_{01}H^{-1}_{11}T_{10}-T_{01}H^{-1}_{11}T_{12}H^{-1}_{22}T_{21}H^{-1}_{11}T_{10}
\end{equation}
By plugging all the matrix into above equation, we have:
\begin{equation}
H_{eff} \approx -\frac{2t^{(dp)}}{\Delta_{CT}} - \frac{4(t^{(dp)})^2}{\Delta_{CT}} (\frac{(t^{(dp)})^2}{u_d} + \frac{2(t^{(dp)})^2}{2\Delta_{CT} + u_p})
\label{Eq:180-pos-afm}
\end{equation}

For $\Delta_{CT}$ << 0 (but we insist $\Delta_{CT}$ < -$u_p$), state $ \frac{1}{\sqrt{2}}(|\uparrow,\downarrow;\uparrow,0;0,\downarrow>-|\uparrow,\downarrow;0,\downarrow;\uparrow,0>)$ and $
\frac{1}{\sqrt{2}}(|\uparrow,0;0,\downarrow;\uparrow,\downarrow>-|0,\downarrow;\uparrow,0;\uparrow,\downarrow>)$ will have lower energy than $\frac{1}{\sqrt{2}}(|\uparrow,0;\uparrow,\downarrow;0,\downarrow>-|0,\downarrow;\uparrow,\downarrow;\uparrow,0>)$, at this time, we arrange the basis as $\{
\frac{1}{\sqrt{2}}(|\uparrow,\downarrow;\uparrow,0;0,\downarrow>-|\uparrow,\downarrow;0,\downarrow;\uparrow,0>), 
\frac{1}{\sqrt{2}}(|\uparrow,0;0,\downarrow;\uparrow,\downarrow>-|0,\downarrow;\uparrow,0;\uparrow,\downarrow>),
|\uparrow,\downarrow;0,0;\uparrow,\downarrow>,
\frac{1}{\sqrt{2}}(|\uparrow,0;\uparrow,\downarrow;0,\downarrow>-|0,\downarrow;\uparrow,\downarrow;\uparrow,0>), 
|\uparrow,\downarrow;\uparrow,\downarrow;0,0>, 
|0,0;\uparrow,\downarrow;\uparrow,\downarrow>, 
\}$ 
and the configuration interaction matrix is:
\begin{equation}
\left[
\begin{array}
{cccccc} 
\Delta_{CT} &0 &-\sqrt{2}t^{(dp)} &t^{(dp)} &-\sqrt{2}t^{(dp)} &0 \\
0 &\Delta_{CT} &-\sqrt{2}t^{(dp)} &t^{(dp)} &0 &-\sqrt{2}t^{(dp)} \\
-\sqrt{2}t^{(dp)} &-\sqrt{2}t^{(dp)} &2\Delta_{CT}+u_p &0 &0 &0 \\
t^{(dp)} &t^{(dp)} &0 &0 &0 &0 \\
-\sqrt{2}t^{(dp)} &0 &0 &0 &u_d &0 \\
0 &-\sqrt{2}t^{(dp)} &0 &0 &0 &u_d \\
\end{array}
\right]
\end{equation}
Here we are interested in $\epsilon \sim \Delta_{CT}$ and take  
\begin{equation}
\begin{split}
H_{00} &= 
\left[
\begin{array}
{cc} 
\Delta_{CT}    &0              \\
0         &\Delta_{CT}         \\
\end{array}
\right] \\
H_{11} &= 
\left[
\begin{array}
{cc} 
2\Delta_{CT} +u_p    &0              \\
0         &0         \\
\end{array}
\right] \\
H_{22} &= 
\left[
\begin{array}
{cc} 
u_d   &0       \\
0     &u_d     \\
\end{array}
\right] \\
\end{split}
\end{equation}

Plugging these terms into {\color{blue}Eq. \ref{Eq:per2}}, we have
\begin{equation}
H_{eff} \approx 
[\Delta_{CT} + (\frac{(t^{(dp)})^2}{\Delta_{CT}} - \frac{2(t^{(dp)})^2}{\Delta_{CT} + u_p})]
\left[
\begin{array}
{cc} 
1    &0              \\
0    &1        \\
\end{array}
\right]
+
(\frac{(t^{(dp)})^2}{\Delta_{CT}} - \frac{2(t^{(dp)})^2}{\Delta_{CT} + u_p})
\left[
\begin{array}
{cc} 
0    &1              \\
1    &0        \\
\end{array}
\right]
\label{Eq:180-neg-afm}
\end{equation}

\subsubsection{B-I-2: $S$ = 1 sector}
There are 9 states in the $S$ = 1 sector:\\
$\{ 
|0,\downarrow;\uparrow,\downarrow;0,\downarrow>,
|\uparrow,\downarrow;0,\downarrow;0,\downarrow>,
|0,\downarrow;0,\downarrow;\uparrow,\downarrow>,
\frac{1}{\sqrt{2}}(|\uparrow,0;\uparrow,\downarrow;0,\downarrow>+|0,\downarrow;\uparrow,\downarrow;\uparrow,0>), 
\frac{1}{\sqrt{2}}(|\uparrow,\downarrow;\uparrow,0;0,\downarrow>+|\uparrow,\downarrow;0,\downarrow;\uparrow,0>), 
\frac{1}{\sqrt{2}}(|\uparrow,0;0,\downarrow;\uparrow,\downarrow>+|0,\downarrow;\uparrow,0;\uparrow,\downarrow>), 
|\uparrow,0;\uparrow,\downarrow;\uparrow,0>,
|\uparrow,\downarrow;\uparrow,0;\uparrow,0>,
|\uparrow,0;\uparrow,0;\uparrow,\downarrow>
\}$ \\
Notice that the first three states, second three states and the last three states are not connected via $t^{(dp)}$, therefore, we can work on a smaller subspace expanded by $\{
|\uparrow,0;\uparrow,\downarrow;\uparrow,0>,
|\uparrow,\downarrow;\uparrow,0;\uparrow,0>,
|\uparrow,0;\uparrow,0;\uparrow,\downarrow>
\}$. 

For $\Delta_{CT}$ >> 0, the configuration interaction matrix in such a basis is:
\begin{equation}
\left[
\begin{array}
{ccc} 
0        &t^{(dp)}     &t^{(dp)}     \\
t^{(dp)} &\Delta_{CT}  &0          \\
t^{(dp)} &0            &\Delta_{CT}     \\
\end{array}
\right] \\
\end{equation}

Taking $\epsilon$ = 0
\begin{equation}
\begin{split}
H_{00} &= 0 \\
H_{11} &= 
\left[
\begin{array}
{cc} 
\Delta_{CT}    &0              \\
0              &\Delta_{CT}    \\
\end{array}
\right] \\
\end{split}
\end{equation}
and making use of {\color{blue}Eq. \ref{Eq:per}}, we have
\begin{equation}
H_{eff} = - \frac{2(t^{(dp)})^2}{\Delta_{CT}}
\label{Eq:180-pos-fm}
\end{equation}

Combining {\color{blue}Eq. \ref{Eq:180-pos-afm}} and {\color{blue}Eq. \ref{Eq:180-pos-fm}}, $J$ is calculated as:
\begin{equation}
\begin{split}
J &= - \frac{2(t^{(dp)})^2}{\Delta_{CT}} + [\frac{2(t^{(dp)})^2}{\Delta_{CT}} + \frac{4(t^{(dp)})^2}{\Delta_{CT}} (\frac{(t^{(dp)})^2}{u_d} + \frac{2(t^{(dp)})^2}{2\Delta_{CT} + u_p})] \\
&= \frac{4(t^{(dp)})^2}{\Delta_{CT}} (\frac{(t^{(dp)})^2}{u_d} + \frac{2(t^{(dp)})^2}{2\Delta_{CT} + u_p}) \\
\end{split}
\label{Eq:180-pos-J}
\end{equation}

For $\Delta_{CT} << 0$ , the basis is rearranged as $\{
|\uparrow,\downarrow;\uparrow,0;\uparrow,0>,
|\uparrow,0;\uparrow,0;\uparrow,\downarrow>,
|\uparrow,0;\uparrow,\downarrow;\uparrow,0>
\}$ and the configuration interaction matrix is
\begin{equation}
\left[
\begin{array}
{ccc} 
\Delta_{CT}   &0            &t^{(dp)}     \\
0             &\Delta_{CT}  &t^{(dp)}     \\
t^{(dp)}      &t^{(dp)}     &0            \\
\end{array}
\right] \\
\end{equation}
Taking $\epsilon$ = $\Delta_{CT}$
\begin{equation}
\begin{split}
H_{00} &= 
\left[
\begin{array}
{cc} 
\Delta_{CT}    &0              \\
0              &\Delta_{CT}    \\
\end{array}
\right] \\
H_{11} &= 0 \\
\end{split}
\end{equation}
and making use of {\color{blue}Eq. \ref{Eq:per}}, we have
\begin{equation}
H_{eff} \approx 
(\Delta_{CT} + \frac{(t^{(dp)})^2}{\Delta_{CT}})
\left[
\begin{array}
{cc} 
1    &0              \\
0    &1        \\
\end{array}
\right]
+
\frac{(t^{(dp)})^2}{\Delta_{CT}}
\left[
\begin{array}
{cc} 
0    &1              \\
1    &0        \\
\end{array}
\right]
\label{Eq:180-neg-fm}
\end{equation}

Since {\color{blue}Eq. \ref{Eq:180-neg-afm}} and {\color{blue}Eq. \ref{Eq:180-neg-fm}} are both 2 $\times$ 2 matrices, we first need to diagonalize it and then $J$ is calculated as:
\begin{equation}
\begin{split}
J &= (\Delta_{CT} + \frac{2(t^{(dp)})^2}{\Delta_{CT}}) - [\Delta_{CT}+ 2(\frac{(t^{(dp)})^2}{\Delta_{CT}} - \frac{2(t^{(dp)})^2}{\Delta_{CT} + u_p}) ] \\
&= \frac{4(t^{(dp)})^2}{\Delta_{CT} + u_p}
\end{split}
\label{Eq:180-neg-J}
\end{equation}
Since $\Delta_{CT} + u_p$ is positive, {\color{blue}Eq. \ref{Eq:180-neg-J}} still gives antiferromagnetic exchange coupling. 

\subsection{B-II: 90$^{\circ}$ geometry}
\subsubsection{B-II-1: $S$ = 0 sector}
There are 10 states in the S = 0 sector: \\
$\{
\frac{1}{\sqrt{2}}(|\uparrow,0;\uparrow,\downarrow;\uparrow,\downarrow;0,\downarrow> - |0,\downarrow;\uparrow,\downarrow;\uparrow,\downarrow;\uparrow,0>),
\frac{1}{\sqrt{2}}(|\uparrow,0;\uparrow,\downarrow;0,\downarrow;\uparrow,\downarrow> -|0,\downarrow;\uparrow,\downarrow;\uparrow,0;\uparrow,\downarrow>),
\frac{1}{\sqrt{2}}(|\uparrow,\downarrow;\uparrow,0;\uparrow,\downarrow;0,\downarrow> -|\uparrow,\downarrow;0,\downarrow;\uparrow,\downarrow;\uparrow,0>),
\frac{1}{\sqrt{2}}(|\uparrow,\downarrow;\uparrow,0;0,\downarrow;\uparrow,\downarrow> - |\uparrow,\downarrow;0,\downarrow;\uparrow,0;\uparrow,\downarrow>),
\frac{1}{\sqrt{2}}(|\uparrow,\downarrow;\uparrow,\downarrow;\uparrow,0;0,\downarrow> - |\uparrow,\downarrow;\uparrow,\downarrow;0,\downarrow;\uparrow,0>),
\frac{1}{\sqrt{2}}(|\uparrow,0;0,\downarrow;\uparrow,\downarrow;\uparrow,\downarrow> - |0,\downarrow;\uparrow,0;\uparrow,\downarrow;\uparrow,\downarrow>),
|\uparrow,\downarrow;\uparrow,\downarrow;\uparrow,\downarrow;0,0>,
|\uparrow,\downarrow;\uparrow,\downarrow;0,0;\uparrow,\downarrow>,
|\uparrow,\downarrow;0,0;\uparrow,\downarrow;\uparrow,\downarrow>,
|0,0;\uparrow,\downarrow;\uparrow,\downarrow;\uparrow,\downarrow>
\}$.
Notice that the first four states are not connected to the last six states via t$_{dp}$, to simplify the discussion, we will work in the subspace spanned by the first four states.

For $\Delta_{CT}$ >> 0, the configuration interaction matrix in the basis $\{\frac{1}{\sqrt{2}}(|\uparrow,0;\uparrow,\downarrow;\uparrow,\downarrow;0,\downarrow> - |0,\downarrow;\uparrow,\downarrow;\uparrow,\downarrow;\uparrow,0>),
\frac{1}{\sqrt{2}}(|\uparrow,0;\uparrow,\downarrow;0,\downarrow;\uparrow,\downarrow> -|0,\downarrow;\uparrow,\downarrow;\uparrow,0;\uparrow,\downarrow>),
\frac{1}{\sqrt{2}}(|\uparrow,\downarrow;\uparrow,0;\uparrow,\downarrow;0,\downarrow> -|\uparrow,\downarrow;0,\downarrow;\uparrow,\downarrow;\uparrow,0>),
\frac{1}{\sqrt{2}}(|\uparrow,\downarrow;\uparrow,0;0,\downarrow;\uparrow,\downarrow> - |\uparrow,\downarrow;0,\downarrow;\uparrow,0;\uparrow,\downarrow>)\}$ is
\begin{equation}
\left[
\begin{array}
{cccc} 
0         &t^{(dp)}                &t^{(dp)}        &0      \\
t^{(dp)}  &\Delta_{CT}-2u_p+5j^{(p)}_H &0           &t^{(dp)} \\
t^{(dp)}  &0         &\Delta_{CT}-2u_p+5j^{(p)}_H   &t^{(dp)} \\
0         &t^{(dp)}  &t^{(dp)} &2\Delta_{CT}-3u_p+9j^{(p)}_H
\end{array}
\right]
\end{equation}
Here we are interested in $\epsilon \sim$ 0 and put
\begin{equation}
\begin{split}
H_{00} &= 0  \\
H_{11} &= 
\left[
\begin{array}
{cc} 
\Delta_{CT}-2u_p+5j^{(p)}_H    &0              \\
0         &\Delta_{CT}-2u_p+5j^{(p)}_H         \\
\end{array}
\right] \\
H_{22} &= 2\Delta_{CT}-3u_p+9j^{(p)}_H \\
\end{split}
\end{equation}
into {\color{blue}Eq. \ref{Eq:per2}}, we have
\begin{equation}
H_{eff} = -\frac{2(t^{(dp)})^2}{(\Delta_{CT} - 2 u_p +5j^{(p)}_H)^2} - \frac{(t^{(dp)})^2}{(\Delta_{CT} - 2 u_p +5j^{(p)}_H)^2}\frac{4(t^{(dp)})^2}{2\Delta_{CT} - 3 u_p +9j^{(p)}_H}
\label{Eq:90-pos-afm}
\end{equation}

For $\Delta_{CT} << 0$ , we rearrange the basis as $\{
\frac{1}{\sqrt{2}}(|\uparrow,\downarrow;\uparrow,0;0,\downarrow;\uparrow,\downarrow> - |\uparrow,\downarrow;0,\downarrow;\uparrow,0;\uparrow,\downarrow>),
\frac{1}{\sqrt{2}}(|\uparrow,0;\uparrow,\downarrow;0,\downarrow;\uparrow,\downarrow> -|0,\downarrow;\uparrow,\downarrow;\uparrow,0;\uparrow,\downarrow>),
\frac{1}{\sqrt{2}}(|\uparrow,\downarrow;\uparrow,0;\uparrow,\downarrow;0,\downarrow> -|\uparrow,\downarrow;0,\downarrow;\uparrow,\downarrow;\uparrow,0>),
\frac{1}{\sqrt{2}}(|\uparrow,0;\uparrow,\downarrow;\uparrow,\downarrow;0,\downarrow> - |0,\downarrow;\uparrow,\downarrow;\uparrow,\downarrow;\uparrow,0>)\}$
and the corresponding configuration interaction matrix is
\begin{equation}
\left[
\begin{array}
{cccc} 
2\Delta_{CT}-3u_p+9j^{(p)}_H           &t^{(dp)}  &t^{(dp)} &0  \\
t^{(dp)}  &\Delta_{CT}-2u_p+5j^{(p)}_H &0         &t^{(dp)} \\
t^{(dp)}  &0     &\Delta_{CT}-2u_p+5j^{(p)}_H     &t^{(dp)} \\
0         &t^{(dp)}                    &t^{(dp)}  &0        \\
\end{array}
\right]
\end{equation}
Here we are interested in $\epsilon \sim (2\Delta_{CT}-3u_p+9j^{(p)}_H)$  and put
\begin{equation}
\begin{split}
H_{00} &= 2\Delta_{CT}-3u_p+9j^{(p)}_H  \\
H_{11} &= 
\left[
\begin{array}
{cc} 
\Delta_{CT}-2u_p+5j^{(p)}_H    &0              \\
0         &\Delta_{CT}-2u_p+5j^{(p)}_H         \\
\end{array}
\right] \\
H_{22} &= 0 \\
\end{split}
\end{equation}
into {\color{blue}Eq. \ref{Eq:per2}}, we have
\begin{equation}
H_{eff} = (2\Delta_{CT}-3u_p+9j^{(p)}_H) + \frac{2(t^{(dp)})^2}{\Delta_{CT} - u_p + 4 j^{(p)}_H} + \frac{(t^{(dp)})^2}{(\Delta_{CT} - u_p + 4 j^{(p)}_H)^2}\frac{4(t^{(dp)})^2}{2\Delta_{CT} - 3u_p + 9 j^{(p)}_H}
\label{Eq:90-neg-afm}
\end{equation}

\subsubsection{B-II-2: $S$ = 1 sector}
There are 18 states in the S = 1 sector: \\
$\{
|0,\downarrow;\uparrow,\downarrow;\uparrow,\downarrow;0,\downarrow>,
|0,\downarrow;\uparrow,\downarrow;0,\downarrow;\uparrow,\downarrow>,
|\uparrow,\downarrow;0,\downarrow;\uparrow,\downarrow;0,\downarrow>,
|\uparrow,\downarrow;0,\downarrow;0,\downarrow;\uparrow,\downarrow>,
|\uparrow,\downarrow;\uparrow,\downarrow;0,\downarrow;0,\downarrow>,
|0,\downarrow;0,\downarrow;\uparrow,\downarrow;\uparrow,\downarrow>,
\frac{1}{\sqrt{2}}(|\uparrow,0;\uparrow,\downarrow;\uparrow,\downarrow;0,\downarrow> + |0,\downarrow;\uparrow,\downarrow;\uparrow,\downarrow;\uparrow,0>),
\frac{1}{\sqrt{2}}(|\uparrow,0;\uparrow,\downarrow;0,\downarrow;\uparrow,\downarrow> +|0,\downarrow;\uparrow,\downarrow;\uparrow,0;\uparrow,\downarrow>),
\frac{1}{\sqrt{2}}(|\uparrow,\downarrow;\uparrow,0;\uparrow,\downarrow;0,\downarrow> +|\uparrow,\downarrow;0,\downarrow;\uparrow,\downarrow;\uparrow,0>),
\frac{1}{\sqrt{2}}(|\uparrow,\downarrow;\uparrow,0;0,\downarrow;\uparrow,\downarrow> + |\uparrow,\downarrow;0,\downarrow;\uparrow,0;\uparrow,\downarrow>),
\frac{1}{\sqrt{2}}(|\uparrow,\downarrow;\uparrow,\downarrow;\uparrow,0;0,\downarrow> + |\uparrow,\downarrow;\uparrow,\downarrow;0,\downarrow;\uparrow,0>),
\frac{1}{\sqrt{2}}(|\uparrow,0;0,\downarrow;\uparrow,\downarrow;\uparrow,\downarrow> + |0,\downarrow;\uparrow,0;\uparrow,\downarrow;\uparrow,\downarrow>),
|\uparrow,0;\uparrow,\downarrow;\uparrow,\downarrow;\uparrow,0>,
|\uparrow,0;\uparrow,\downarrow;\uparrow,0;\uparrow,\downarrow>,
|\uparrow,\downarrow;\uparrow,0;\uparrow,\downarrow;\uparrow,0>,
|\uparrow,\downarrow;\uparrow,0;\uparrow,0;\uparrow,\downarrow>,
|\uparrow,\downarrow;\uparrow,\downarrow;\uparrow,0;\uparrow,0>,
|\uparrow,0;\uparrow,0;\uparrow,\downarrow;\uparrow,\downarrow>
\}$\\
Notice that the first six states, the second six states and the last six states are not connected via t$_{dp}$, in other words, we can work on either subspace. Here we will work on $\{
|\uparrow,0;\uparrow,\downarrow;\uparrow,\downarrow;\uparrow,0>,
|\uparrow,0;\uparrow,\downarrow;\uparrow,0;\uparrow,\downarrow>,
|\uparrow,\downarrow;\uparrow,0;\uparrow,\downarrow;\uparrow,0>,
|\uparrow,\downarrow;\uparrow,0;\uparrow,0;\uparrow,\downarrow>,
|\uparrow,\downarrow;\uparrow,\downarrow;\uparrow,0;\uparrow,0>,
|\uparrow,0;\uparrow,0;\uparrow,\downarrow;\uparrow,\downarrow>
\}$. What's more, notice that the first four states are also not connected with the last two via $t^{(dp)}$, for simplicity, we will work on the subspace spanned by the first four: $\{
|\uparrow,0;\uparrow,\downarrow;\uparrow,\downarrow;\uparrow,0>,
|\uparrow,0;\uparrow,\downarrow;\uparrow,0;\uparrow,\downarrow>,
|\uparrow,\downarrow;\uparrow,0;\uparrow,\downarrow;\uparrow,0>,
|\uparrow,\downarrow;\uparrow,0;\uparrow,0;\uparrow,\downarrow>
\}$.

With $\Delta_{CT}$ >> 0, the configuration interaction matrix under $\{
|\uparrow,0;\uparrow,\downarrow;\uparrow,\downarrow;\uparrow,0>,
|\uparrow,0;\uparrow,\downarrow;\uparrow,0;\uparrow,\downarrow>,
|\uparrow,\downarrow;\uparrow,0;\uparrow,\downarrow;\uparrow,0>,
|\uparrow,\downarrow;\uparrow,0;\uparrow,0;\uparrow,\downarrow>
\}$ is
\begin{equation}
\left[
\begin{array}
{cccc} 
0         &t^{(dp)}                    &t^{(dp)}  &0        \\
t^{(dp)}  &\Delta_{CT}-2u_p+5j^{(p)}_H &0         &t^{(dp)} \\
t^{(dp)}  &0         &\Delta_{CT}-2u_p+5j^{(p)}_H &t^{(dp)} \\
0         &t^{(dp)}  &t^{(dp)}  &2\Delta_{CT}-3u_p+7j^{(p)}_H  \\
\end{array}
\right]
\end{equation} 
Here we are interested in $\epsilon \sim 0$ and put  
\begin{equation}
\begin{split}
H_{00} &= 0  \\
H_{11} &= 
\left[
\begin{array}
{cc} 
\Delta_{CT}-2u_p+5j^{(p)}_H    &0              \\
0         &\Delta_{CT}-2u_p+5j^{(p)}_H        \\
\end{array}
\right] \\
H_{22} &= 2\Delta_{CT}-3u_p+7j^{(p)}_H  \\
\end{split}
\end{equation}
into {\color{blue}Eq. \ref{Eq:per2}}, we have
\begin{equation}
H_{eff} = -\frac{2(t^{(dp)})^2}{(\Delta_{CT} - 2 u_p +5j^{(p)}_H)^2} - \frac{(t^{(dp)})^2}{(\Delta_{CT} - 2 u_p +5j^{(p)}_H)^2}\frac{4(t^{(dp)})^2}{2\Delta_{CT} - 3 u_p +7j^{(p)}_H}
\label{Eq:90-pos-fm}
\end{equation}

Combining {\color{blue}Eq. \ref{Eq:90-pos-afm}} and {\color{blue}Eq. \ref{Eq:90-pos-fm}}, $J$ is calculated as:
\begin{equation}
J = -\frac{4(t^{(dp)})^2}{(\Delta_{CT} - 2 u_p +5j^{(p)}_H)^2}[\frac{(t^{(dp)})^2}{(2\Delta_{CT} - 3 u_p +8j^{(p)}_H) - j^{(p)}_H} - \frac{(t^{(dp)})^2}{(2\Delta_{CT} - 3 u_p +8j^{(p)}_H) + j^{(p)}_H}] 
\label{Eq:90-pos-J}
\end{equation}

With $\Delta_{CT} << 0$ , the basis is rearranged as $\{
|\uparrow,\downarrow;\uparrow,0;\uparrow,0;\uparrow,\downarrow>,
|\uparrow,0;\uparrow,\downarrow;\uparrow,0;\uparrow,\downarrow>,
|\uparrow,\downarrow;\uparrow,0;\uparrow,\downarrow;\uparrow,0>,
|\uparrow,0;\uparrow,\downarrow;\uparrow,\downarrow;\uparrow,0>
\}$ and the configuration interaction matrix now becomes
\begin{equation}
\left[
\begin{array}
{cccc} 
2\Delta_{CT}-3u_p+7j^{(p)}_H &t^{(dp)}   &t^{(dp)}  &0        \\
t^{(dp)}    &\Delta_{CT}-2u_p+5j^{(p)}_H &0         &t^{(dp)} \\
t^{(dp)}    &0        &\Delta_{CT}-2u_p+5j^{(p)}_H  &t^{(dp)} \\
0           &t^{(dp)} &t^{(dp)}          &0                   \\
\end{array}
\right]
\end{equation} 
Here we are interested in $\epsilon \sim (2\Delta_{CT}-3u_p+7j^{(p)}_H)$ and put  
\begin{equation}
\begin{split}
H_{00} &= 2\Delta_{CT}-3u_p+7j^{(p)}_H   \\
H_{11} &= 
\left[
\begin{array}
{cc} 
\Delta_{CT}-2u_p+5j^{(p)}_H    &0              \\
0         &\Delta_{CT}-2u_p+5j^{(p)}_H        \\
\end{array}
\right] \\
H_{22} &= 0 \\
\end{split}
\end{equation}
into {\color{blue}Eq. \ref{Eq:per2}}, we have
\begin{equation}
H_{eff} = (2\Delta_{CT}-3u_p+7j^{(p)}_H) + \frac{2(t^{(dp)})^2}{\Delta_{CT} - u_p + 2 j^{(p)}_H} + \frac{(t^{(dp)})^2}{(\Delta_{CT} - u_p + 2 j^{(p)}_H)^2}\frac{4(t^{(dp)})^2}{2\Delta_{CT} - 3u_p + 7 j^{(p)}_H}
\label{Eq:90-neg-fm}
\end{equation}

Combining {\color{blue}Eq. \ref{Eq:90-neg-afm}} and {\color{blue}Eq. \ref{Eq:90-neg-fm}}, we have:
\begin{equation}
J = -2 j^{(p)}_H
\label{Eq: 90-neg-J}
\end{equation}
where higher orders are ignored.

\subsection{B-III: Further reduction for $\Delta_{CT} >> 0$ }
For $\Delta_{CT} >> 0$, $p$ is much deeper than $d$ orbital, under such circumstances, $p$ orbital can be treated as uncorrelated one with $u_p$ = 0 and $j^{(p)}_H$ = 0 (but the last $j^{(p)}_H$ term of {\color{blue}Eq-(\ref{eq:Hd})} is kept). So $\Delta_{CT} = \epsilon_d - \epsilon_p + u_d = \epsilon_{dp} + u_d$ and {\color{blue}Eq. \ref{Eq:180-pos-J}}, {\color{blue}Eq. \ref{Eq:90-pos-J}} are simplified to:
\begin{equation}
\begin{split}
J &= \frac{4(t^{(dp)})^2}{(\epsilon_{dp} + u_d)^2}(\frac{1}{u_d} + \frac{1}{\epsilon_{dp} + u_d}) \\
J &=-\frac{4(t^{(dp)})^2}{(\epsilon_{dp} + u_d)^2}(\frac{1}{2(\epsilon_{dp}+u_d)-j^{(p)}_H} - \frac{1}{2(\epsilon_{dp}+u_d)+j^{(p)}_H}) \\
\end{split}
\end{equation}
in accordance with the results of E. Koch \cite{Koch2012}.

\newpage
\section{APPENDIX C: Multi-band extension}
\begin{figure*}[h]
\centering
\includegraphics[width=8cm]{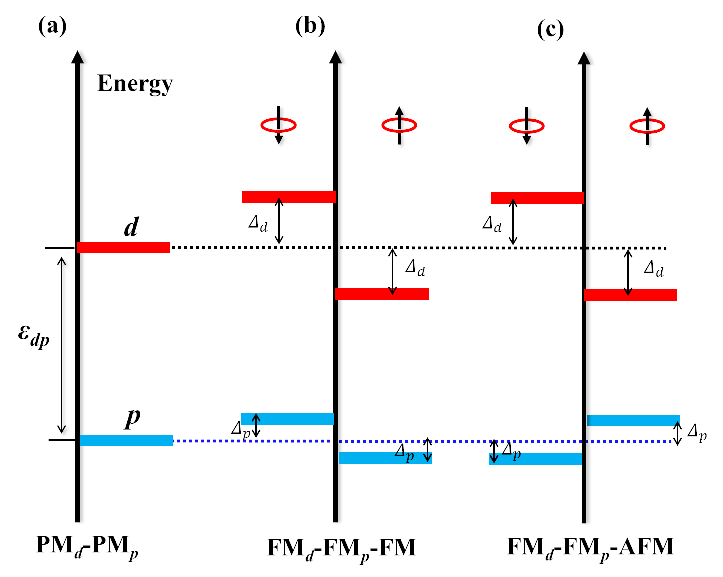}
\caption{(a) Energy level in the PM phase. (b) Energy level in the $\rm FM_d-FM_p-FM$ phase. (c) Energy level in the $\rm FM_d-FM_p-AFM$ phase.  
\label{Fig:AFM-FM} }
\end{figure*}

For simplicity, here we consider the case $t^{(d)}$ = 0, $t^{(p)}$ = 0. In the PM phase, $p$ orbital lies $\varepsilon_{dp}$ below $d$ orbital as shown in {\color{blue}Fig. \ref{Fig:AFM-FM}(a)}. Suppose now both $d$ and $p$ subsystem are FM because of strong interaction, there are two possible configurations for the whole system: FM$_d$-FM$_p$-FM in {\color{blue}Fig. \ref{Fig:AFM-FM}(b)} and FM$_d$-FM$_p$-AFM in {\color{blue}Fig. \ref{Fig:AFM-FM}(c)}. For FM$_d$-FM$_p$-AFM, since $p$ orbital has opposite polarization with respect to $d$ orbital, the spin up electron will have higher energy than spin down electron. This is the difference between {\color{blue}Fig. \ref{Fig:AFM-FM}(b)} and {\color{blue}Fig. \ref{Fig:AFM-FM}(c)}. In the following, we will study the following two cases: 1) $d$ is half filled and 2) $d$ is empty. Since the two $p$ orbitals are degenerate as well as the two $d$ orbitals, in the following, we will only consider one $p$ and one $d$ orbital.

\subsection{C-I: $d$ orbital is half filled}
Now we include $t^{(dp)}$. By fixing $N_e$ to 3, the energy of  FM$_d$-FM$_p$-FM and FM$_d$-FM$_p$-AFM is the sum of the lowest three energy level. For FM$_d$-FM$_p$-FM, we have:
\begin{equation}
E(FM_d-FM_p-FM) = -[\varepsilon_{dp}-(\Delta_d - \Delta_p)] - \frac{[\varepsilon_{dp}+(\Delta_d - \Delta_p)]+\sqrt{[\varepsilon_{dp}+(\Delta_d - \Delta_p)]^2 + 4(t^{(dp)})^2}}{2}
\end{equation}
where $\Delta_p$, $\Delta_d$ is half of the Zeeman splitting due to the intrinsic magnetic order.
On the other hand, the energy for FM$_d$-FM$_p$-AFM is:
\begin{equation}
E(FM_d-FM_p-AFM) = -[\varepsilon_{dp}-(\Delta_d + \Delta_p)] - \frac{[\varepsilon_{dp}+(\Delta_d + \Delta_p)]+\sqrt{[\varepsilon_{dp}+(\Delta_d + \Delta_p)]^2 + 4(t^{(dp)})^2}}{2}
\end{equation}

The energy difference between FM$_d$-FM$_p$-FM and FM$_d$-FM$_p$-AFM can be seen as a function of $\Delta_p$:
\begin{equation}
\begin{split}
f_1(\Delta_p) &= E(\rm FM_d-FM_p-FM) - E(\rm FM_d-FM_p-AFM)   \\
&=-\Delta_p - \frac{\sqrt{[\varepsilon_{dp}+(\Delta_d - \Delta_p)]^2 + 4(t^{(dp)})^2}}{2} + \frac{\sqrt{[\varepsilon_{dp}+(\Delta_d + \Delta_p)]^2 + 4(t^{(dp)})^2}}{2} \\
\end{split}
\end{equation}
When $\Delta_p$ = 0, $f_1(\Delta_p)$ = 0 as expected. Taking derivative on $\Delta_p$, we have:
\begin{equation}
\frac{d}{d \Delta_p} f_1(\Delta_p) = -1 + \frac{1}{2}\frac{1}{\sqrt{1 + (\frac{2t^{(dp)}}{\varepsilon_{dp}+(\Delta_d - \Delta_p)})^2}} + \frac{1}{2}\frac{1}{\sqrt{1 + (\frac{2t^{(dp)}}{\varepsilon_{dp}+(\Delta_d + \Delta_p)})^2}} < -1 + \frac{1}{2} + \frac{1}{2} = 0
\end{equation}
Combing with the fact that $f_1(\Delta_p = 0)$ = 0, we have $E(FM_d-FM_p-FM)$ < $E(FM_d-FM_p-AFM)$ when $\Delta_p \neq 0$. This is the main reason why the ground state for $\Delta_{CT}$ closing to 0 is FM$_d$-FM$_p$-FM in {\color{blue}Fig. \ref{fig:mf}(c)}, not FM$_d$-FM$_p$-AFM.

\subsection{C-II: $d$ orbital is empty}
Now we consider the case when $d$ is empty, in other words, we have
$N_e$ = 2. The energy of FM$_d$-FM$_p$-FM and FM$_d$-FM$_p$-AFM is the sum of the lowest two energy level. For FM$_d$-FM$_p$-FM, we have:
\begin{equation}
\begin{split}
E(FM_d-FM_p-FM) &=\frac{-[\varepsilon_{dp}-(\Delta_d - \Delta_p)]-\sqrt{[\varepsilon_{dp}-(\Delta_d - \Delta_p)]^2 + 4(t^{(dp)})^2}}{2} \\&+\frac{-[\varepsilon_{dp}+(\Delta_d - \Delta_p)]-\sqrt{[\varepsilon_{dp}+(\Delta_d - \Delta_p)]^2 + 4(t^{(dp)})^2}}{2} \\
\end{split}
\end{equation}
The energy for FM$_d$-FM$_p$-AFM is:
\begin{equation}
\begin{split}
E(FM_d-FM_p-AFM) &= \frac{-[\varepsilon_{dp}-(\Delta_d + \Delta_p)]-\sqrt{[\varepsilon_{dp}-(\Delta_d + \Delta_p)]^2 + 4(t^{(dp)})^2}}{2} \\
&+ \frac{-[\varepsilon_{dp}+(\Delta_d + \Delta_p)]-\sqrt{[\varepsilon_{dp}+(\Delta_d + \Delta_p)]^2 + 4(t^{(dp)})^2}}{2} \\
\end{split}
\end{equation}

Then $f_2(\Delta_p)$ is given by:
\begin{equation}
\begin{split}
f_2(\Delta_p) &= E(FM_d-FM_p-FM) - E(FM_d-FM_p-AFM)   \\
&= \frac{\sqrt{[\varepsilon_{dp}-(\Delta_d + \Delta_p)]^2 + 4(t^{(dp)})^2}}{2} + \frac{\sqrt{[\varepsilon_{dp}+(\Delta_d + \Delta_p)]^2 + 4(t^{(dp)})^2}}{2} \\
&- \frac{\sqrt{[\varepsilon_{dp}-(\Delta_d - \Delta_p)]^2 + 4(t^{(dp)})^2}}{2} -\frac{\sqrt{[\varepsilon_{dp}+(\Delta_d - \Delta_p)]^2 + 4(t^{(dp)})^2}}{2} \\
\end{split}
\end{equation}
When $\Delta_p$ = 0, $f_2(\Delta_p)$ = 0 as expected. Taking derivative on $\Delta_p$, we have:
\begin{equation}
\begin{split}
\frac{d}{d \Delta_p} f_2(\Delta_p) &= \frac{1}{2}(\frac{1}{\sqrt{1+(\frac{2t^{(dp)}}{\varepsilon_{dp}+(\Delta_d + \Delta_p)}})^2} -\frac{1}{\sqrt{1+(\frac{2t^{(dp)}}{\varepsilon_{dp}+(\Delta_d - \Delta_p)}})^2}) \\
&+\frac{1}{2}(\frac{1}{\sqrt{1+(\frac{2t^{(dp)}}{\varepsilon_{dp}-(\Delta_d - \Delta_p)}})^2} -\frac{1}{\sqrt{1+(\frac{2t^{(dp)}}{\varepsilon_{dp}-(\Delta_d + \Delta_p)}})^2})\\
\end{split}
\end{equation}
Since we have
\begin{equation}
\begin{split}
\varepsilon_{dp}+(\Delta_d + \Delta_p) > \varepsilon_{dp}+(\Delta_d - \Delta_p) \\
\varepsilon_{dp}-(\Delta_d - \Delta_p) > \varepsilon_{dp}-(\Delta_d + \Delta_p)
\end{split}
\end{equation}
we have $\frac{d}{d \Delta_p} f_2(\Delta_p) > 0$. Together with the fact that $f_2(\Delta_p = 0)$ = 0, we have $E(FM_d-FM_p-FM)$ > $E( FM_d-FM_p-AFM)$ when $\Delta_p \neq 0$. Therefore, FM$_d$-FM$_p$-AFM will be the preferred ground state for $\Delta_{CT}$ closing to 0 when $d$ is empty.

\subsection{C-III: One $d$ orbital is half-filled and one $d$ orbital is empty}
We now stack the above two systems together (there is no communication between these two systems). The $j^{(p)}_H$ and $j^{(d)}_H$ align the spins on the two $d$ and two $p$ orbitals along the same direction. Now we ask which phase is more stable, FM$_d$-FM$_p$-FM or FM$_d$-FM$_p$-AFM ? This is equivalent to calculate the sign of the following energy:
\begin{equation}
f(\Delta_p) = f_1(\Delta_p) + f_2(\Delta_p)
\end{equation}
Obviously, we have $f(\Delta_p = 0)$ = 0. The derivative on $\Delta_p$ is:
\begin{equation}
\frac{d}{d \Delta_p} f(\Delta_p) = -1 + \frac{1}{\sqrt{1 + (\frac{2t^{(dp)}}{\varepsilon_{dp}+(\Delta_d - \Delta_p)})^2}} + \frac{1}{\sqrt{1 + (\frac{2t^{(dp)}}{\varepsilon_{dp}+(\Delta_d + \Delta_p)})^2}} + \frac{1}{2}(\frac{1}{\sqrt{1+(\frac{2t^{(dp)}}{\varepsilon_{dp}-(\Delta_d - \Delta_p)}})^2} -\frac{1}{\sqrt{1+(\frac{2t^{(dp)}}{\varepsilon_{dp}-(\Delta_d + \Delta_p)}})^2})
\end{equation}
As long as $t^{(dp)}$ is not too large, we will have $\frac{d}{d \Delta_p} f(\Delta_p) > 0$. Therefore, FM$_d$-FM$_p$-AFM will be the preferred phase for partial filled $d$ shell in general.

\newpage
\section{APPENDIX D: Model parameters calculation}

\begin{figure*}
\includegraphics[width=16cm]{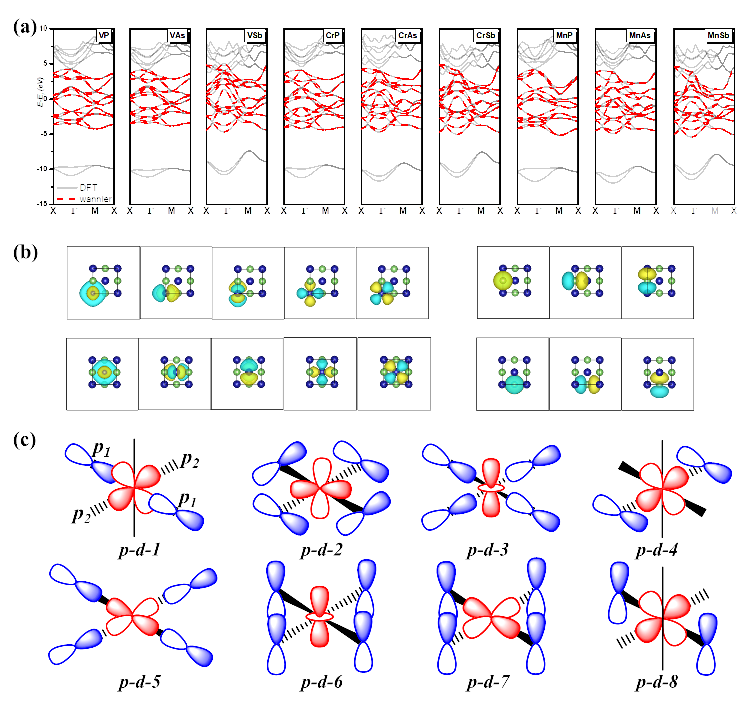}
\caption{(a) PBE level PM band structure of all 9 M$L$ monolayers. (b) The representative maximally localized WFs of PM CrAs monolayer. (c) Eight symmetry-allowed $p-d$ hopping channel. The dashed/bold wedge represents bonds above/below the paper plane, following the Natta projection in stereochemistry. The $p$ orbitals above (below) the paper plane is labelled as $p_1$ ($p_2$).}
\label{Fig:PBE}
\end{figure*}

\subsection{D-I: Single-particle part}
The single-particle parameters such as onsite energy ($\varepsilon_{d}, \varepsilon_{p}$) and hopping energy ($t^{(dp)}, t^{(p)}, t^{(p)}$) in {\color{blue}Eq-(\ref{eq:Full-H})} can be obtained by downfolding the full Hamiltonian into the $\{ d, p \}$ subspace in Wannier90 package\cite{Wanpac} with Perdew-Burke-Ernzerhof (PBE) functional \cite{PBE}. Explicitly, the downfolding process also allows us to obtain the following matrix element:
\begin{equation}
H_{\alpha\beta}(R) = <\phi_{0,\alpha}|\hat{H}|\phi_{R,\beta}>
\end{equation}
where $|\phi_{0,\alpha}>$ is the maximally localized Wannier function $\alpha$ in home cell (index as 0) and  $|\phi_{R,\beta}>$ the maximally localized Wannier function (MLWF) $\beta$ in cell R. When R=0, $\alpha=\beta$, the above matrix element orbital energy, otherwise we obtain the hopping energy. 

The PM band structure and the Wannier fitted one is shown in {\color{blue}Fig. \ref{Fig:PBE}(a)}. The corresponding 16 maximally localized WFs for CrAs monolayer are displayed in {\color{blue}Fig. \ref{Fig:PBE}(b)}. The excellent agreement of the band structures and the small spreading of MLWFs indicates the downfolding process is quite good, which lays the foundation of cRPA calculation below. In cRPA calculation, 72 bands are used with $5 \times 5 \times 1$ meshes for Brillouin zone integration.

The space group of this lattice structure is $\rm P\frac{4}{n}mm$, all the  $p-d$ hybridizations can be classified into eight types (labelled as $p-d-i$, i = 1, 2, ..., 8) displayed in {\color{blue}Fig. \ref{Fig:PBE}(c)}.  
Explicitly, the $p-d-1$ composes of ${d_{xz}}$ and ${p_{1x}}$ as well as the equivalent ${d_{yz}}$ and ${p_{2y}}$. The $p$ orbitals of $L$ above and below M plane are marked as $p_1$ and $p_2$ correspondingly. And the orbital contribution to other $p-d$ channels is listed in {\color{blue}Tab. \ref{Tab:tdp}}. When there is no buckling in this structure, the point group becomes ${\rm D_{4h}}$. Only $p-d-2$, $p-d-3$, $p-d-5$ and $p-d-8$ exist in ${\rm D_{4h}}$ with $p-d-5$ the strongest, which can be as large as 1.3 eV in cuprates \cite{CupJ}. When the structure is buckled, there are no $90^{\circ}$ and $180^{\circ}$ M-L-M angles and $p-d-1$, $p-d-4$, $p-d-6$, $p-d-7$ appear.

\begin{table}
\setlength{\belowcaptionskip}{10pt}
\centering
\caption{ Exchange coupling strength and Curie temperature for different system}
\begin{tabular}{p{28mm}<{\centering}p{20mm}<{\centering}p{20mm}<{\centering}}
\toprule
Channel  &   $d$     & $p$ \\
\midrule
$p-d-1$     & ${d_{xz}}$ (${d_{yz}}$)   & ${p_{1x}}$ (${p_{2y}}$)  \\
$p-d-2$     & ${d_{xy}}$                & ${p_{1y} + p_{2x}}$      \\
$p-d-3$     & ${d_{z^2}}$               & ${p_{1x} + p_{2y}}$      \\
$p-d-4$     & ${d_{xz}}$ (${d_{yz}}$)   & ${p_{2x}}$ (${p_{1y}}$)  \\
$p-d-5$     & ${d_{x^2-y^2}}$           & ${p_{1x} + p_{2y}}$      \\
$p-d-6$     & ${d_{z^2}}$               & ${p_{1z} + p_{2z}}$      \\
$p-d-7$     & ${d_{x^2-y^2}}$           & ${p_{1z} + p_{2z}}$      \\
$p-d-8$     & ${d_{xz}}$ (${d_{yz}}$)   & ${p_{1z}}$ (${p_{2z}}$)  \\
\bottomrule
\end{tabular}
\label{Tab:tdp}
\end{table}

\subsection{D-II: Interaction term}
In solid, the Coulomb potential in solid is screened by electronic polarizability  and is thus renormalized. The constrained random phase approximation (cRPA) provides a systematic first-principles technique for the construction of low-energy Hamiltonians where the interaction part is calculated \cite{cRPA1, cRPA2, cRPA3}. 

Clearly there is a degree of freedom in choosing target subspace. Different subspace orbitals will give different interaction parameters ($u, u', j_H$), at the same time, the number of parameters are also 
changed as well as the onsite energy and hopping strength. In MHI, it is often enough to choose part of M $d$ orbitals as the target space. In our case here, we need to treat {\color{blue}Eq-(\ref{eq:Full-H})}, so both M $d$ and $L$ $p$ orbitals are chosen to be target subspace.

After calculation, the value of the screened interaction ($W^r$) between local orbitals is expressed as 4-index interaction matrix \cite{Roekeghem2016}:
\begin{equation}
U^S_{m_1m_2m_3m_4}(\omega) 
= <\phi_{m_1} \phi_{m_2}|W^r(\omega)|\phi_{m_3} \phi_{m_4}> 
\end{equation}
where $\omega$ is the frequency which describes the dynamical effect (only $\omega$ = 0 is used here),  the $|\phi_{m}>$ is the localized Wannier orbitals, and $S$ is added for specifying the angular symmetry of the localized Wannier orbitals. Here we take cubic angular harmonics approximation so S is cubic. Most matrix elements are of the order of 0.1 eV or less, except for 2-index reduced interaction matrices \cite{Roekeghem2016}. Furthermore, 2-index matrix can be further simplified as scalars which is used in model Hamiltonian like {\color{blue}Eq-(\ref{eq:Full-H})}. For either $d$ or $p$ shell, there are three independent intra-shell values:
\begin{equation}
\begin{split}
u &= \frac{1}{N} \sum_{m=1}^{N} U^{cubic}_{mmmm}  \\
u^{\prime} &= \frac{1}{N(N-1)} \sum_{m \neq n}^{N} U^{cubic}_{mnmn} \\
j_H &= \frac{1}{N(N-1)} \sum_{m \neq n}^{N} U^{cubic}_{mnnm} \\
\end{split}
\end{equation}
where N = 5 (3) for $d$ ($p$) shell. As for the inter-shell Coulomb interaction, only the density-density interaction $u_{dp}$ is calculated as all the other interaction terms are nearly zero. 
\begin{equation}
u_{dp} = \frac{1}{15} \sum_{m, n} U^{cubic}_{dmpn}
\end{equation}

\newpage
\section{Appendix E: Exchange coupling strength}
\begin{figure*}
\includegraphics[width=10cm]{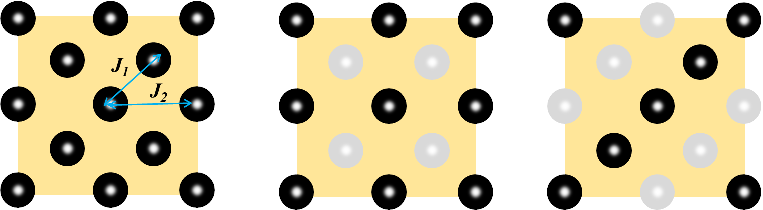}
\caption{Schematic view of three ordered magnetic states. The black/white solid sphere represents up/down spin magnetic moment, respectively.}
\label{Fig:J-map-1}
\end{figure*}

Here we apply energy mapping method to calculate $J$. In this method, the total energies of different spin configuration are calculated and the exchange interactions are fitted to the energies of different spin configuration. For example, to map $J_1$, $J_2$ out, three symmetric magnetic orders are considered in a ${2 \times 2 \times 1}$ supercell as shown in {\color{blue}Fig. \ref{Fig:J-map-1})}: FM with magnetic ordering momentum \textbf{q}=(0, 0), checkboard AFM (c-AFM) with \textbf{q}=$(\pi, \pi)$ and strip AFM (s-AFM) with \textbf{q}=$(0, \pi)$. The spin model used here is the Heisenberg model:
\begin{small}
\begin{equation}
H=\sum_{i,j}J_{ij}\vec{S_i}\cdot\vec{S_j}
\end{equation}
\end{small}
\noindent where summation is over $J_1$ and $J_2$ as defined in {\color{blue}Fig. 4(d)}. And the energy for these three magnetic orders are calculated as:
\begin{equation}
\begin{split}
E_{FM}  &= E_0 + 16J_1 S^2 + 16J_2 S^2 \\
E_{c-AFM} &= E_0 - 16J_1 S^2 + 16J_2 S^2 \\
E_{s-AFM} &= E_0 - 16J_2 S^2 \\
\end{split}
\end{equation}
\noindent where $E_0$ is a reference energy. These energies are obtained via HSE06 functional. By taking $E_{FM}$, $E_{c-AFM}$ and $E_{s-AFM}$ into above equations, $J_1$ and $J_2$ can be obtained for a given $S$. 

When $S$ is in the classical limit, energy mapping method has shown success in FM CHI and MHI as $S$ is easy to define there \cite{PASP, Li2014}. This is not the case in FNCHI as $L$ is also polarized. However the net magnetic moment on $L$ is quite small (see {\color{blue}Fig. \ref{Fig:Functional}(e)}), which makes it inappropriate to denote an integer magnetic moment attached to $L$. What's more, to compensate the holes on As, extra electrons are back donated to Cr, so the magnetic moment on Cr is a slightly larger than that without back-donation. So here we make the following simplification: by treating a Cr and its nearby four As as a whole, it is possible to denote an half-integer $S$ = 3/2 to it. In this way, the $J$ can be evaluated by energy mapping method. 

The calculated relative energy of $E_{FM}$, $E_{c-AFM}$ and $E_{s-AFM}$ for CrAs monolayer is 0, +0.511 and +0.632 eV, respectively. By taking $S$ = 3/2, the obtained $J_1$ and $J_2$ in CrAs monolayer is -56.8 and -39.7 meV. 

\newpage
\section{APPENDIX F: Scaling of exchange interaction strength with respect to distance}
\begin{figure*}
\includegraphics[width=16cm]{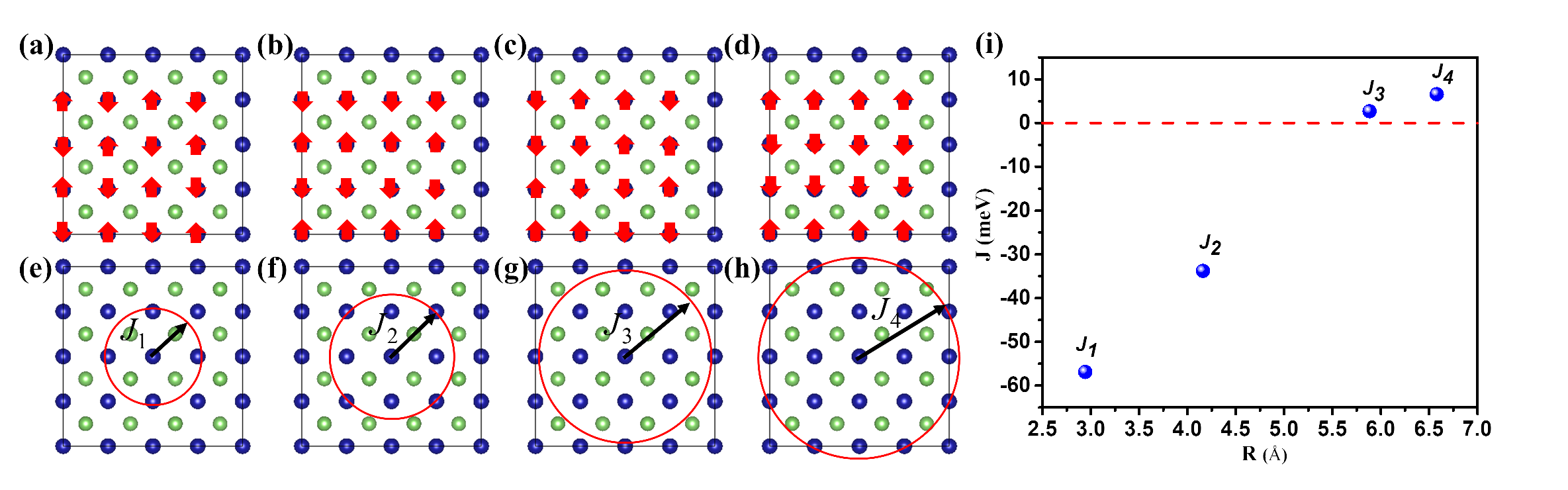}
\caption{(a) Checkboard AFM configuration. (b) Strip AFM configuration. (c) Bi-strip AFM configuration. (d) Anti-stripe AFM configuration. The up and down red arrow in (a)-(d) represents the orientation of local magnetic moments. Real space illustration of range of (e) $J_1$, (f) $J_2$, (g) $J_3$ and (h) $J_4$. (f) Strength of $J$ with respect to the interaction distance.}
\label{Fig:J-map-2}
\end{figure*}
Since the spreading of MMO provides a natural length scale, it is interesting to see whether $J$ have different behavior inside and outside this length scale. To see this, we use energy mapping method on a 4 $\times$ 4 supercell and calculate $J$ up to $J_4$ as shown in {\color{blue}Fig. \ref{Fig:J-map-2}(a)-(h)}. As it is difficult for HSE06 to handle such a large supercell, here we use SCAN functional for a compromise. The energies of the different magnetic orders are:
\begin{equation}
\begin{split}
E_{FM} &= E_0 + 32J_1 S^2 + 32J_2 S^2 + 32J_3 S^2 + 64J_4 S^2  \\
E_{c-AFM} &= E_0 - 32J_1 S^2 + 32J_2 S^2 + 32J_3 S^2 - 64J_4 S^2 \\
E_{s-AFM} &= E_0 - 32J_2 S^2 + 32J_3 S^2 \\
E_{bis-AFM} &= E_0 - 32J_3 S^2 \\
E_{antis-AFM} &= E_0 + 16J_1 S^2 - 32J_4 S^2 \\
\end{split}
\end{equation}

The result is displayed in {\color{blue}Fig. \ref{Fig:J-map-2}(i)}. It is clear that $J$ experiences a sharp decrease from $J_2$ to $J_3$. From this result, it is reasonable to use $J_1$ and $J_2$ obtained in Appendix E in evaluating $T_c$ in the main text. For $J_1$ and $J_2$, their interaction lengths are within the spreading of MMOs, while for $J_3$ and $J_4$, they are beyond the spreading of MMOs. Therefore, MMO indeed provides a natural length scale for $J$. 

\end{appendices}

\end{document}